\documentclass[12pt]{article}

\usepackage{graphicx}
\usepackage{subfigure}

\usepackage{enumitem}
\setlist{leftmargin=5.5mm}

\setlist[enumerate, 1]{
  leftmargin=\parindent,
  parsep=.5\parsep,
  itemsep=0ex
}

\setlist[itemize, 1]{
  leftmargin=\parindent,
  parsep=.5\parsep,
  itemsep=0ex
}

\usepackage{pdfsync}
\usepackage[active]{srcltx}
\usepackage{color}
\usepackage{longtable}
\definecolor{darkblue}{RGB}{25,18,130}

\usepackage[colorlinks]{hyperref}
\hypersetup{urlcolor=blue}

\usepackage{amsmath,amssymb,amsthm}
\usepackage{newtxtext}
\usepackage{newtxmath}
\usepackage{textcomp}
\usepackage{stmaryrd}
\usepackage{proofsteps}
\usepackage{bbm}
\usepackage{float}
\usepackage{tcolorbox}
\usepackage[normalem]{ulem}

\theoremstyle{plain}
\newtheorem{thm}{Theorem}[section]

\newtheorem{prop}[thm]{Proposition}

\newtheorem{cor}{Corollary}

\theoremstyle{definition}
\newtheorem{defn}{Definition}[section]

\theoremstyle{remark}

\newtheorem{exm}{Example}

\def\expmath{Maude}

\newcommand{\ceiling}[1]{\lceil #1 \rceil}
\newcommand{\floor}[1]{\lfloor #1 \rfloor}
\def\trans{\to}

\usepackage{listings}
\lstdefinelanguage{Maude}{%
  morekeywords={%
    mod, endm, fmod, endfm, th, endth, fth, endfth, is,
    protecting,
    sort, sorts, subsort,
    var, vars,
    op, ops,
    eq, ceq, rl, crl,
    view, from, to, endv,
    search
  }%
}
\newcommand{\maudelisting}{%
  \lstset{%
    language=Maude,
    basicstyle=\ttfamily,
    keywordstyle=\bfseries,
    columns=flexible,
    keepspaces=true,
    escapechar=\#
  }%
}

\usepackage{setspace}
\lstnewenvironment{maude}{%
  \spacing{1}
  \maudelisting
}{%
  \endspacing
}

\usepackage{connectives}
\usepackage{xypic}

\newcommand{\co}{\,\colon\;}

\def\MSA{\mbox{$\mathit{MSA}$}}
\def\POA{\mbox{$\mathit{POA}$}}

\usepackage{authblk}

\begin{document}
\title{Computational Modelling for Combinatorial Game
  Strategies}  
\author[$\dagger$]{R\u{a}zvan Diaconescu}

\date{}


\maketitle

\begin{abstract}
We develop a generic computational model that can be used effectively
for establishing the existence of winning strategies for concrete
finite combinatorial games.
Our modelling is (equational) logic-based involving advanced
techniques from algebraic specification, and it can be executed by
equational programming systems such as those from the OBJ-family.
We show how this provides a form of experimental mathematics for
strategy problems involving combinatorial games.
We do this by defining general methods and by illustrating these
with case studies. 
\end{abstract}

\section{Introduction}

The existence of (winning) strategies for finite combinatorial 
games has been established by Zermelo in \cite{zermelo1913} (an
English translation of this paper can be found in
\cite{schwalbe-walker2001}). 
That work is widely considered as pioneering in game theory.
Although Zermelo's original work was confined to the game of chess,
his result holds in general for games between two players that
alternate turns, that always terminate, that lack any chance aspect
(such as dice games).
Moreover, the final result of playing a game is that one of the
players `wins' while the other `loses', whatever that means.
This can be extended with `draw' situations, when neither of the two
players wins or loses.
Zermelo's result is applied in some areas of computing science,
for instance in model checking. 

In our paper we develop a (conditional) equational logic
axiomatisation of strategies for the above-mentioned class of games
that is based on Boolean-valued functions on trees. 
This axiomatisation is abstract in the sense that is independent of
particular games.
We code it as a parameterized/generic functional module in Maude
\cite{maude-book}, a high-performance algebraic specification language
that belongs to the OBJ-family \cite{iobj}. 
The generic character of this axiomatisation and of its coding allows
for a uniform method to obtain very high-level running code that can
be executed (by rewriting) to establish the existence of brute
strategies for concrete particular games.  
For this, we need only to write an executable equational specification
of the respective particular game tree and then use it as an instance 
of the parameter of the generic specification module.

In general, a strategy provides a complete and deterministic plan for
how a player will act in every possible situation in a game.
It specifies exactly what action the player will take at each decision
point, given any information they may have.
A ``winning strategy'' guarantees that the respective player always
ends in a winning position. 
By \emph{brute} (formulations of) strategies we mean those
strategies that are given exhaustively in the sense that all
states of the game are considered and at each state a specific move is
specified.
In general, brute strategies represent huge data, so they have
to be handled by computing systems.
By contrast, \emph{smart} (formulations of) strategies come as
generic and uniform, and can be expressed in a natural language.  
An example of the latter would be ``if possible, color a  cell that
does not neighbour any other already coloured cell'',\footnote{This is
  not a real winning strategy for the game of Example
  \ref{board-game}, for none of the players, and is given here only
  for the purpose to illustrate how natural language can be used to
  formulate smart strategies.
  At this moment we do not want to spoil this problem by revealing the
  real smart winning strategies for the game.}
while for the same game (which is the game of Example \ref{board-game}
in Section \ref{gamesex}) the implementation of the smart strategy as
an exhaustive list of cases counts as a brute strategy. 
The existence of smart strategies implies the existence of brute
ones, but the converse is far from being straightforward.
So, if we establish that brute winning strategies do not exist then
there is no hope for a smart winning strategy.
Otherwise, we can start thinking towards the formulation of a
smart winning strategy, an enterprise that is usually highly
non-trivial as it requires insight, creativity, imagination.
However, all these can be cultivated to a great extent and,
furthermore, computational experiments can provide substantial support
to this process.

The smart-brute concepts of strategies are informal rather than
mathematical concepts.
In Section \ref{strategies-as-subtrees} we will introduce also a
mathematical concept of strategy, which corresponds to the brute
strategies in the sense discussed above.
For smart strategies there is no such mathematical concept. 

Our work is originally motivated by experimental mathematics, which
means that we actually use its results for establishing when brute
strategies exist and eventually `extract' patterns, a kind of
information that may raise mathematical insight leading first to
smart formulation of strategies and then to mathematical proofs
validating them. 
However, as we will see with one of our examples, our results can be
applied beyond the experimental mathematics formulation. 

\paragraph{Prerequisites.}
Fundamentally, our choice of Maude as the concrete programming
language in which our equational specifications are coded can be seen
as arbitrary because our work involves the most standard form of
equational logic that is used in algebraic specification, plus a
common form of parameterised specifications.
In this sense, we can do it with many other executable algebraic
specification frameworks.
On the other hand, Maude's rewriting engine is superior to its
competitors, and this aspect plays a big role in fighting the inherent
complexity of some of our applications/examples.

In Section \ref{poa-sec} we will also resort to `non-deterministic
rewriting', a logic-based computing paradigm that goes beyond
equational logic. 
Up to our knowledge only two languages realise this paradigm, namely
Maude and CafeOBJ \cite{caferep, cafefun}. 

It is evident that here is not a suitable place to introduce these
specification and programming paradigms, nor their mathematical
foundations, in a non-trivial self-contained manner.
For that, the reader may study some general algebraic specification
literature (such as \cite{sannella-tarlecki-book} or more specific
ones such as \cite{maude-book,caferep}).
On the other hand, Maude notation is so close to the standard
mathematical notation in many-sorted equational logic, that for the
reader with enough theoretical background, Maude code can be read
almost without any recourse to other materials.
However, when some notations are Maude-specific we provide the
necessary explanations.

In any case, some familiarity with algebraic specification theory is
an asset for studying this work.  

\subsection{The structure of the paper}

\begin{enumerate}[leftmargin=1.5em]

\item We develop our own form of Zermelo's theorem that is based on
  Boolean-valued functions on game trees.
  This is directly suitable for logic-based computational modelling. 
  But before doing that, we illustrate the kind of games that are
  subject of this result by few concrete examples.
  Although all these fall within the scope of our theory, there are
  several significant differences between them.

\item We turn our proof of Zermelo's theorem into a parameterised
  algebraic specification, as an abstract axiomatisation in the
  equational logic of \emph{many-sorted algebra} (abbreviated \MSA).
  The first part of Section \ref{imp-sec} will be dedicated to a very
  succint presentation of some basic concepts of this logic. 

\item The parameter of the generic specification can be instantiated
  to specifications of trees of concrete games.
  We illustrate how this can be done with the concrete examples of the
  games presented at the beginning of the first section.
  One of them will be presented in greater detail, while in the case
  of the others, although they are fully developed, we will discuss
  only their main ideas.
  The full code of our examples is available in indicated
  repositories. 
  Moreover, we explain the general mechanism of this instantiation
  process, which has solid foundations in category theory.
  We will also include a discussion about programming techniques to
  fight the complexity of the game trees. 

\item Finally, we arrive at the experimental mathematics side of our 
  work.
  We discuss runs of the concrete equational programs obtained by
  instantiating the generic specification based on our
  interpretation and proof of Zermelo's theorem.  
  Consequently, we obtain data that reveals the positions where there
  are winning strategies for each player.
  Moreover, we also present some additional programming techniques
  that support the finding of patterns leading to crucial insight
  into the problems, further leading to smart formulations of
  winning strategies and to mathematical arguments validating them.
  One of them is based on running through winning positions, while the
  other generalises the main equational program for computations of
  traces of winning strategies. 

\end{enumerate}

\section{Zermelo's theorem in a strategies-as-subtrees perspective}

In this section we first provide some examples of combinatorial
games. 
Then we define strategies as subtrees of the game trees.
Finally, in this context, we prove our own version of Zermelo's
theorem.  

\subsection{Some game examples}
\label{gamesex}

\begin{exm}[A heaps of tokens game]
  \label{heap-game}
The following problem was proposed in December 2021 in the
mathematical magazine \emph{K\"{o}MaL} in the section ``Advanced
Problems in  Mathematics''.
\begin{quote}
Rebecca and Benny play the following game: there are two heaps of
tokens, and they take turns to pick some tokens from them. 
The winner of the game is the player who takes away the last token.
If the number of tokens in the two heaps are $A$ and $B$ at a given
moment, the player whose turn it is can take away a number of tokens
that is a multiple of $A$ or a multiple of $B$ from one of the heaps.
Rebecca plays first. 

Find those pairs of integers $(k,n)$, for which Benny has a winning
strategy, if the initial number of tokens is $k$ in the first heap and
$n$ in the second heap. 
\end{quote}
\end{exm}

\begin{exm}[A board colouring game]
  \label{board-game}
This is an example of a concrete game in which each player performs
its own kind of moves.
I saw it in a problem set on combinatorics proposed to Romanian
students preparing for the \emph{Junior Balkan Mathematical Olympiad}.
\begin{quote}
Benny and Rebecca colour the cells of an $n \times n$ board in
blue and red as follow.
First, Benny colours a $2 \times 2$ square in blue, then Rebecca
  colours a single cell in red, and this alternation of colourings
  gets repeated until a next colouring step cannot be performed anymore.
  When this happens, all remaining uncoloured cells get coloured in
  red by default. 
If at the end there are more blue than red squares, then Benny
  wins.
  But if there are more red than blue squares then Rebecca wins. 
  The question is: do Benny or Rebecca have a winning strategy?
\end{quote}
Below is a sample of a playing a game:
\begin{figure}[thbp]
    \centering
    \subfigure{\includegraphics[width=0.16\textwidth]{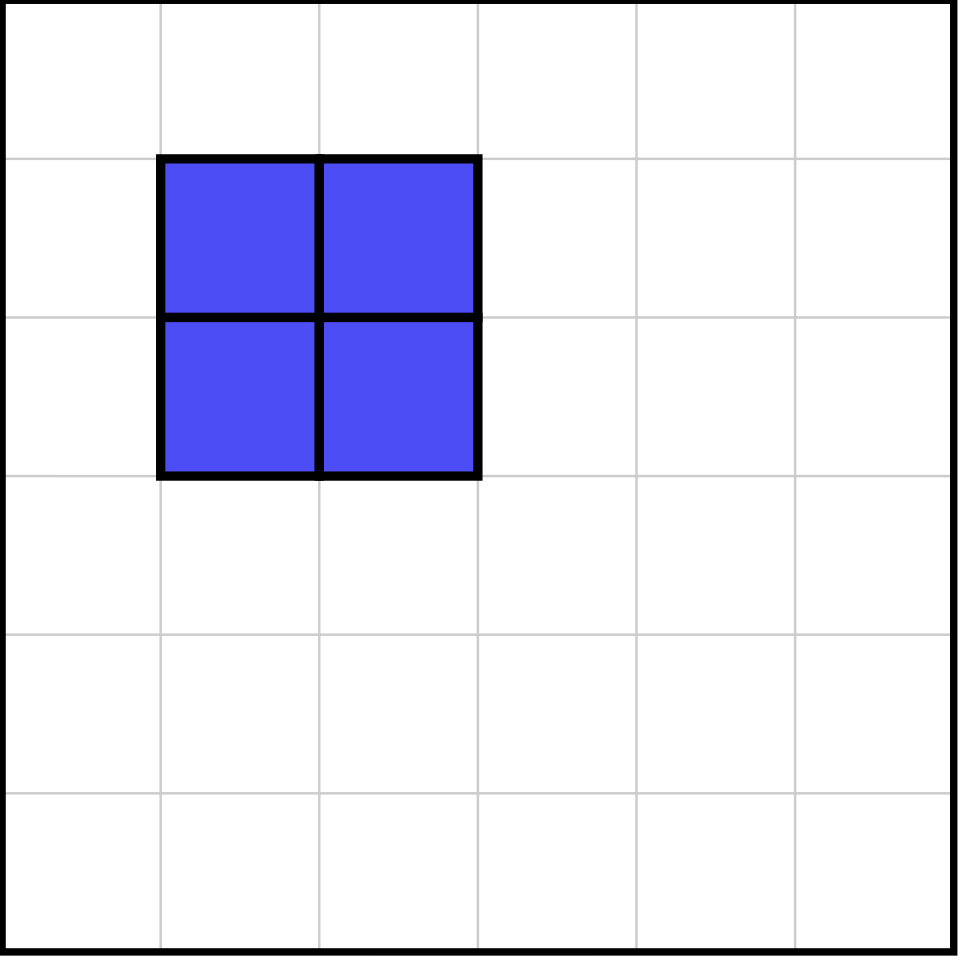}}
    \hspace{1em}
    \subfigure{\includegraphics[width=0.16\textwidth]{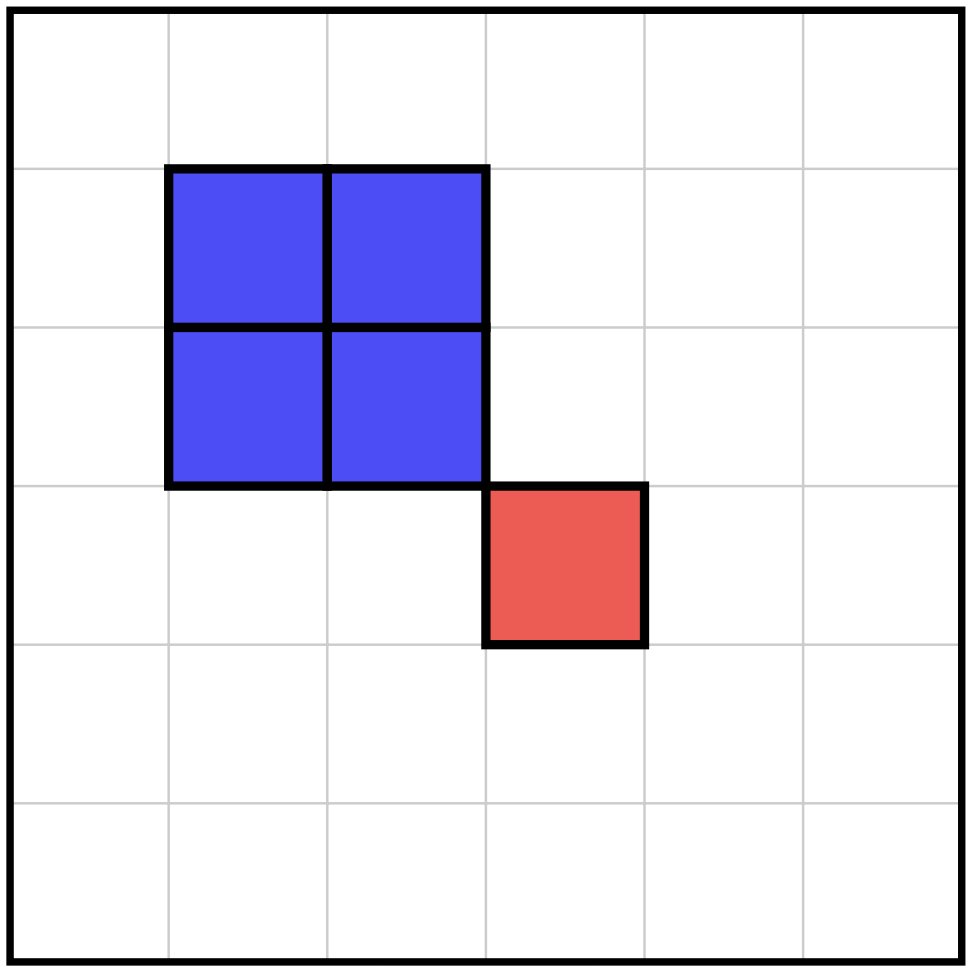}}
    \hspace{1em}
    \subfigure{\includegraphics[width=0.16\textwidth]{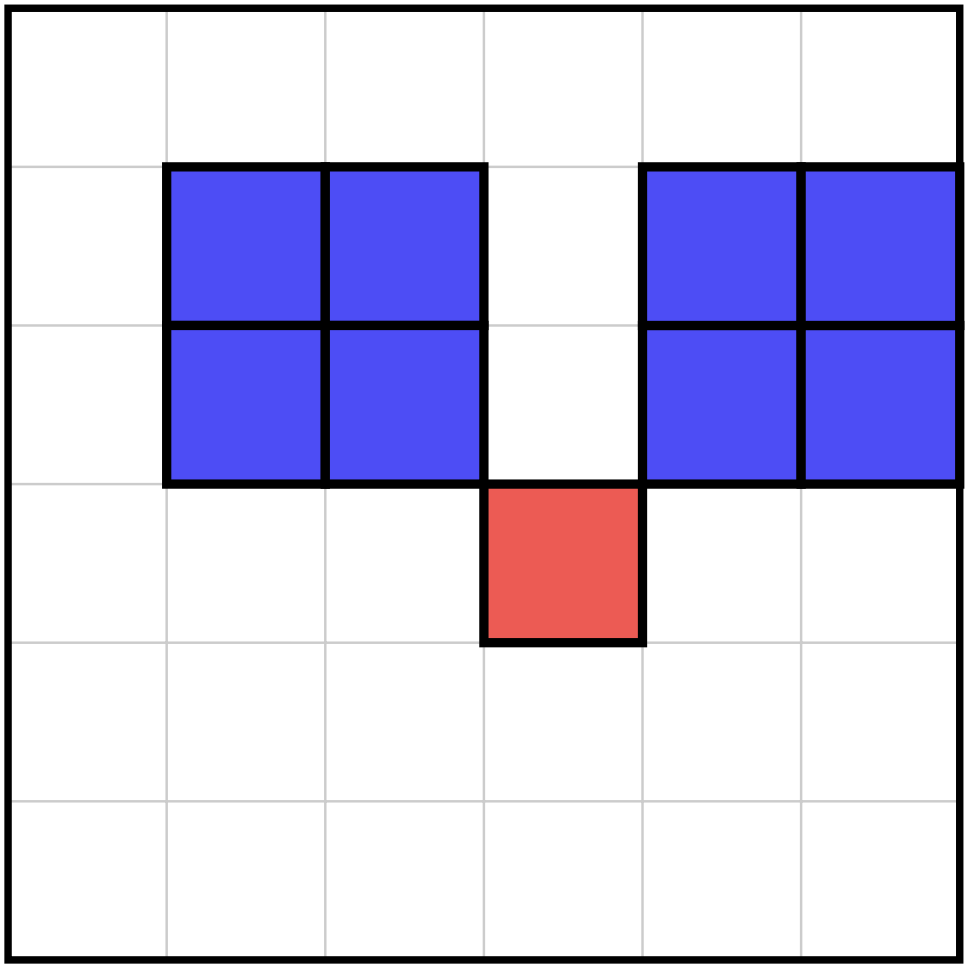}}
    \hspace{1em}
    \subfigure{\includegraphics[width=0.16\textwidth]{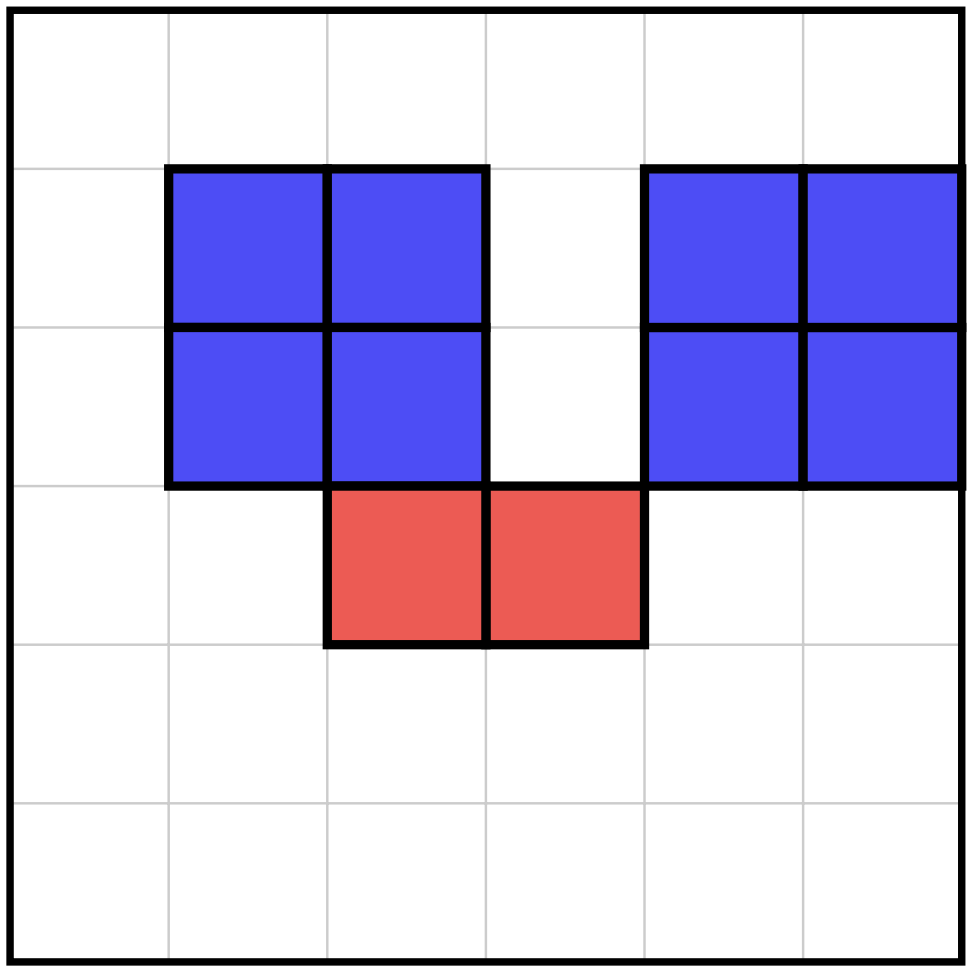}}
    \hspace{1em}
    \subfigure{\includegraphics[width=0.16\textwidth]{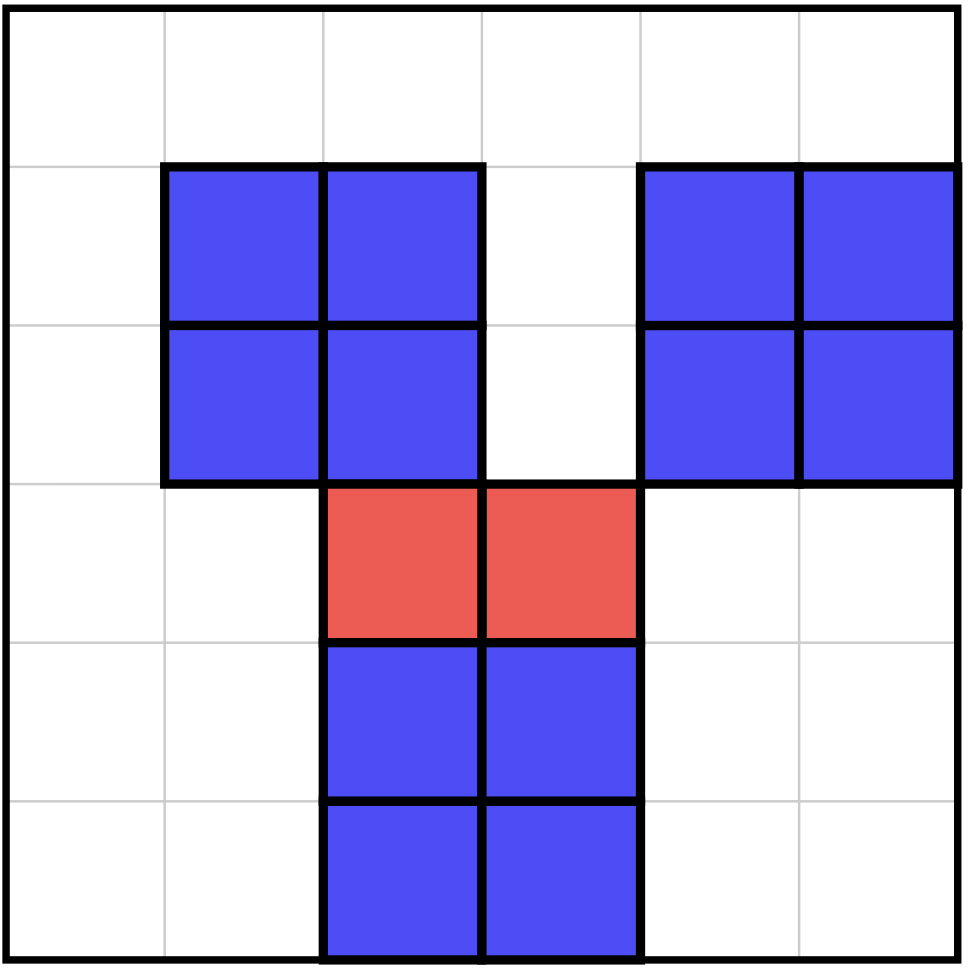}} 
\end{figure}

\begin{figure}[h]
    \centering
    \subfigure{\includegraphics[width=0.16\textwidth]{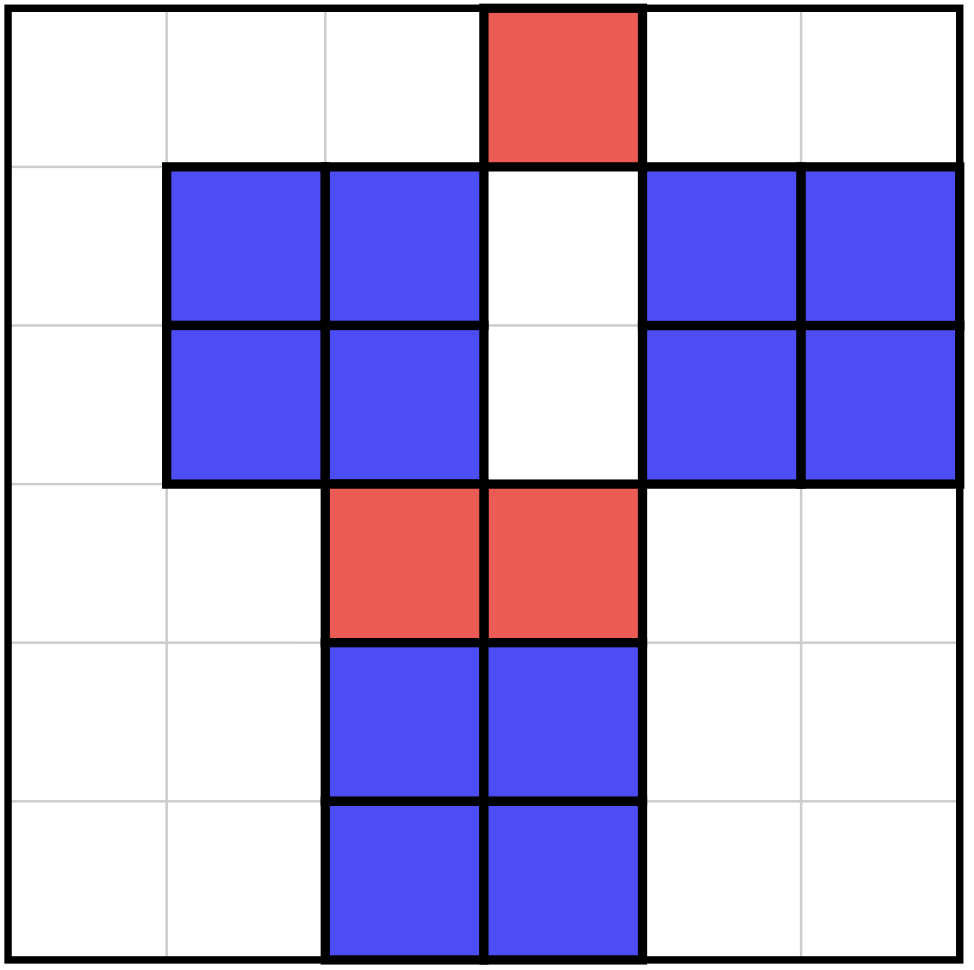}}
    \hspace{1em}
    \subfigure{\includegraphics[width=0.16\textwidth]{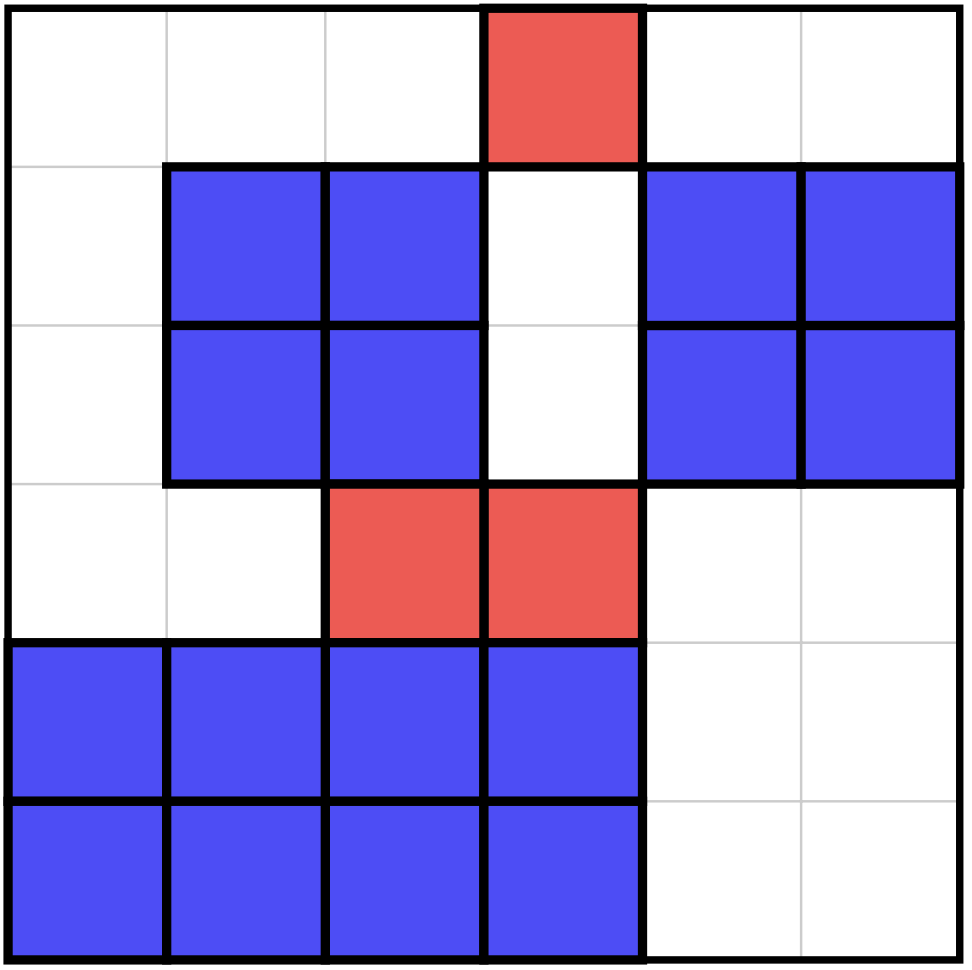}}
    \hspace{1em}
    \subfigure{\includegraphics[width=0.16\textwidth]{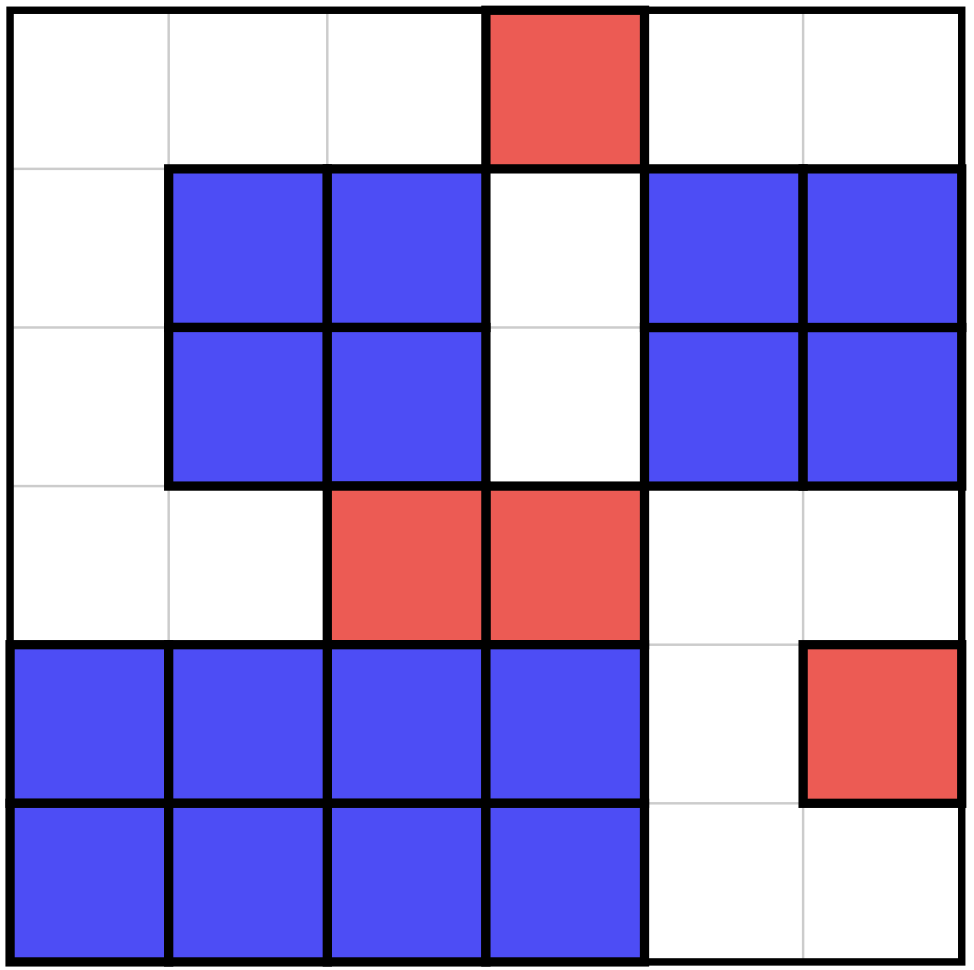}}
    \hspace{1em}
    \subfigure{\includegraphics[width=0.16\textwidth]{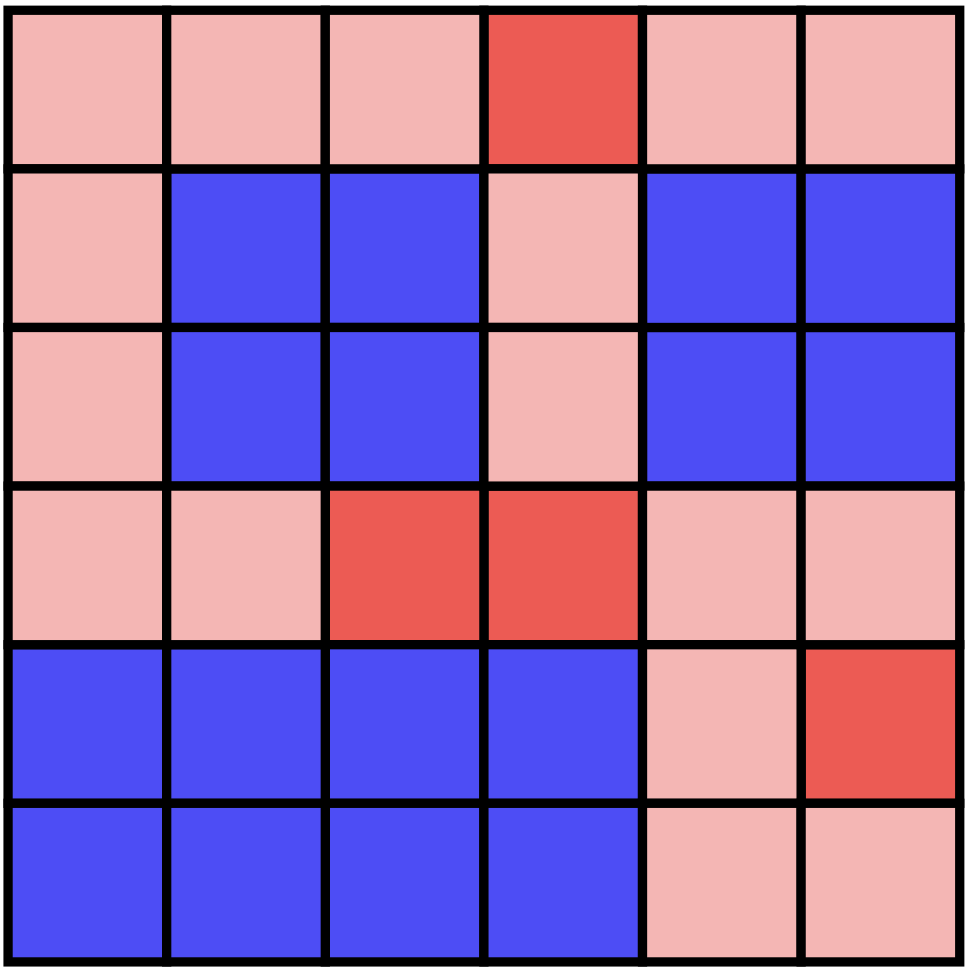}}
  \caption{A board colouring play -- red wins.}
    \label{fig:board-colouring-play}
\end{figure}
  
\end{exm}



\begin{exm}[A transition systems game]
  \label{bisim-exm}
This example comes from computer science theory.
A \emph{transition system} \cite{keller76} is a binary relation,
written $(S,T)$, where $S$ is a set (of ``states'') and $T \subseteq S
\times S$ (called a ``transition'' relation).
A pair $(s_1, s_2) \in T$ is called a \emph{transition}, and may be
denoted as \( \xymatrix{s_1 \rto & s_2} \).
Two simple particular examples of finite transition systems are the
following ones:
\[
  \xymatrix @R-2ex {
    (S,T): & a \rto & b \rto & c \\
    & & v \drto & \\
    (S',T'): & u \drto \urto & & z \\
     & & w \urto & 
    }
\]
This is the simplest version of the concept, as there are more
sophisticated forms of transition systems, such as the \emph{labelled
  transition systems} where each transition has a certain label.
Our approach extends easily to labelled transition systems, but for
reasons of simplicity of presentation we stick with the simpler
version.

A \emph{path} in a transition system $(S,T)$ is a sequence of transitions
\[
  \xymatrix{
    s_0 \rto & s_1 \rto & \ldots \rto & s_n \rto & \ldots.  
}
\]
The game is as follows:
\begin{quote}
Given two transition systems $(S,T), (S',T')$, and two states, $s_0 \in
S$ and $s'_0 \in S'$, Benny and Rebecca play the following game, in
rounds, that builds two paths, one in each of the two transition
systems. 
During each round, Benny starts by picking a transition that extends
one of the two paths, which means that after Benny moves one of the two
paths becomes one state longer.
Then, Rebecca extends the shorter path, so after her move the equality
between the length of the two paths is restored.
Benny wins when Rebecca cannot move, while Rebecca wins when Benny
cannot move or when there exists $k < n$ such that
$(s_n, s'_n)= (s_k, s'_k)$, where $n$ is the current length of the two
paths.   
\end{quote}
\end{exm}

Let us note the following significant differences between these games:
\begin{itemize}[leftmargin=2em]

\item According to the established terminology in game theory, the
  heaps game is \emph{impartial} in the sense that both players
  perform the same kind of moves. 
  This is not the case with the board and the transition systems
  games, where each player performs its own kind of moves.
  Hence, the latter games are called \emph{partisan}. 

\item While the heaps and the transition systems games are win-lose
  ones, in general the board game is a win-lose-draw game (it is a
  win-lose game when the size of the board is odd). 

\end{itemize}
We strongly encourage the reader, that before reading more into this
paper, to try to solve by himself the strategy questions of the heaps
and the board colouring games. 
Only then he can really understand their degree of difficulty.
One of the benefits of the experimental mathematics method emerging
from our work is that it may effectively help to crack such
problems. 

\subsection{Strategies as subtrees}
\label{strategies-as-subtrees}

Our approach is based on representing games by trees, in the sense of
graph theory.
For any graph $G$ let $V(G)$ denote the set of its vertices (nodes)
and $E(G)$ the set of its edges.
A \emph{tree} is an acyclic connected graph.
An \emph{out-tree} (or \emph{arborescence}) is a directed rooted
(i.e. it has a designated vertex called `root') tree having a single
(directed) path from the root to each vertex. 
A \emph{subtree} $S$ of a tree $T$ is a tree whose nodes and edges
belong to $T$.
A \emph{leaf} is any vertex without out-going edges. 

\begin{figure}[H]
\begin{center}
\scalebox{.8}{
\includegraphics{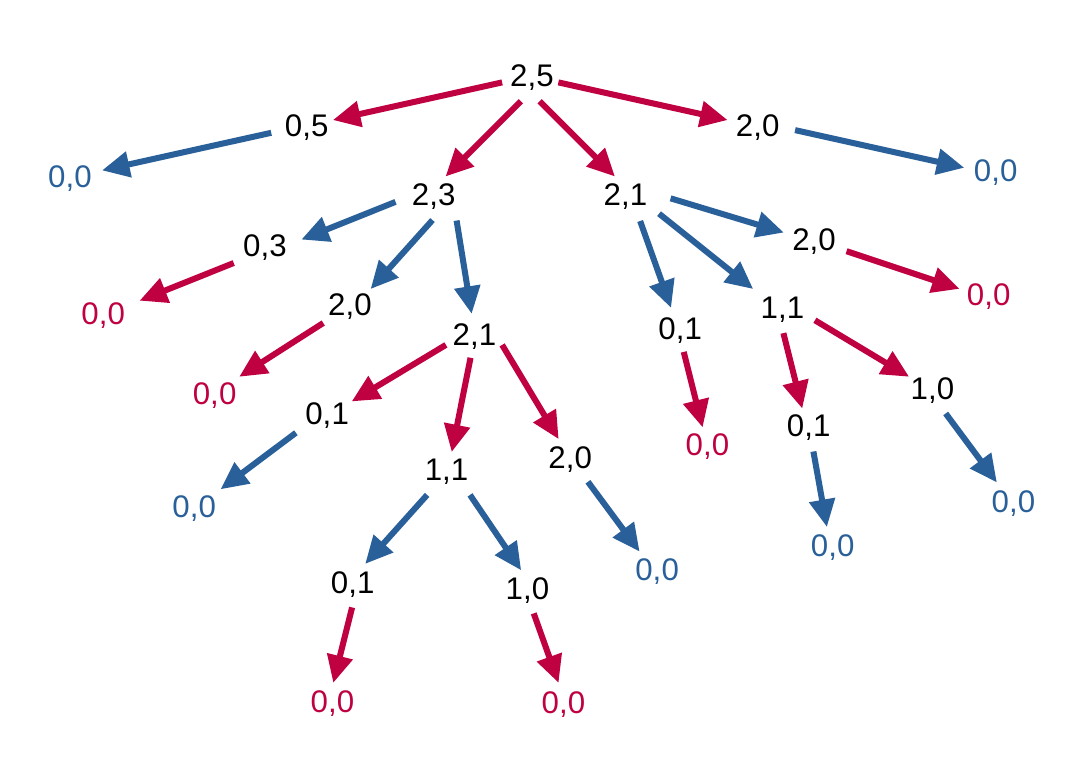}
}
\caption{The tree of the game of Example \ref{heap-game} with the
  heaps containing initially $2$ and $5$ tokens, respectively.}
\label{a812-1fig}
\end{center}
\end{figure}

\paragraph{Game trees.}
The representation of combinatorial games as finite out-trees (which
is standard for \emph{extensive-form} games\footnote{Extensive-form
  games are games allowing
  for the explicit representation of the sequencing of players'
  possible moves, their choice at every decision point, the
  information each player has about the opponent's moves when they
  make a decision, and their payoffs for all possible outcomes. A
  formal definition can be found in \cite{hart1992}.}) goes as
follows: 
\begin{itemize}[leftmargin=1.5em]

\item The vertices represent configurations (or states) of the game,
  whatever these are.

\item The edges represent all possible moves in the game. 

\item The edges are coloured according to who makes the respective
  move.
  So, there are two colours for the edges.
  For instance, in the case of the games of our examples we can use
  blue (for Benny) and red (for Rebecca). 

\item Each sequence of moves leads to a different node in the tree.
  This means we can have the same game configuration at various
  different nodes.
  Or, if we do not like this, we can consider that the game
  configurations also carry all history that produced that particular
  configuration. 

\item Each leaf is coloured in blue, red, or black.
  The colour of a leaf shows who won at the respective leaf (terminal
  configuration).
  Black represents a draw. 

\end{itemize}
Figure \ref{a812-1fig} shows the game tree for the game of Example
\ref{heap-game}, when the initial configuration is $(2,5)$. 
The following conventions will be useful.
If a game is played by $B$ and $R$, then for $X \in \{ B,R \}$, an
\emph{$X$-edge} is an edge in the game tree that corresponds to a move
by $X$.
An \emph{$X$-node} is a node such that every outgoing edge is an
$X$-edge.  

\paragraph{Strategies as subtrees.}
Informally, $R$ has a winning strategy when she can win regardless of
how $B$ moves. 
There can be several ways to win depending on $R$'s decisions.
This idea is captured by the following definition. 

\begin{defn}\label{strategy-dfn1}
Consider a game with two players, $B$ and $R$, that alternate their
moves.
Let $G$ be a game tree.
Then an \emph{$R$-strategy}\footnote{$B$-strategies get a similar
  definition.} is a \emph{subtree} $S$ of $G$ that satisfies
the following properties: 
\begin{enumerate}[leftmargin=1.5em]

\item $S$ shares the same root with $G$,

\item for each node of $S$ \emph{all} outgoing $B$-edges in $G$
  belong to $S$, 

\item for each node of $S$ \emph{exactly one} outgoing $R$-edge in
  $G$ belongs to $S$.
  
\end{enumerate}
When all leaves of $S$ represent winning situations for $R$ (according
to the specifications of the actual game) we say that $S$ is a
\emph{winning $R$-strategy}. 
\end{defn}

By the terminology discussed in the introduction, we can classify
these strategies as being brute.
In principle, strategies in this sense can be very numerous.  
However, in the case of the particular game showed in Figure
\ref{a812-1fig} the only winning strategy for Rebecca is given by the
subtree that starts with the move $2,5 \to 2,3$ and from $2,1$ moves
to $1,1$.

In the context of the transition systems game of Example
\ref{bisim-exm}, the existence of a winning strategy for Rebecca is
equivalent to the states $s_0$ and $s'_0$ being \emph{bisimilar}, a
situation defined by the existence of a \emph{bisimulation} relation
$R \subseteq S \times S'$ such that $(s, s') \in R$ (see
\cite{stirling99}). 
$R$ is a bisimulation when for any $(u,u') \in R$, for any
$\xymatrix{u \rto & v} \in T$ there exists
$\xymatrix{u' \rto & v'} \in T'$ such that $(v,v') \in R$ and
viceversa (see \cite{milner89}).
\[
  \xymatrix @+1ex {
    u \ar@{.}[r]^R \ar[d]_T & u' \ar[d]^{T'} \\
    v \ar@{.}[r]_R & v'
  }
\]
In the case of the two particular concrete transition systems of
Example \ref{bisim-exm}, we have that $a$ and $u$ are bisimilar while
$a$ and $v$ are not.\footnote{We have actually checked these facts by
  programs that will be discussed below in the paper.} 

\subsection{No strategy means that the opponent has a strategy}

Definition \ref{strategy-dfn1} is very general as it does not care
about what are the actual moves in the game and also does not depend
on any concrete formulation of a winning criteria either.
In this very general context, it is possible to prove the following
version of the main result from \cite{zermelo1913}. 

\begin{thm}\label{strategy-thm1}
  $B$ and $R$ play a game that has the following characteristics:
  \begin{enumerate}[leftmargin=1.5em]

  \item The players alternate their moves.

  \item The game always terminates.

  \item In each terminal situation there are three mutually exclusive
    possibilities: either one and only one of the two players is
    declared winner\footnote{This counts as two possibilities.} or
    else the game is declared a draw. 
  \end{enumerate}
Then the absence of a winning $R$-strategy implies the existence of a
non-losing $B$-strategy. 
\end{thm}

\begin{proof}
  Let $G$ be the game tree.
  By recursion we define the following Boolean-valued function $w \co
  V(G) \to \{ 0,1 \}$:
  \begin{equation}\label{w-def1}
w(x) =
\begin{cases}
  1, & x \text{ leaf and $R$ wins} \\
  0, & x \text{ leaf and $R$ does \emph{not} win} \\
  \bigwedge_{xy \in E(G)}\limits w(y), & x \ B\text{-node} \\
   \bigvee\limits_{xy \in E(G)} w(y), & x \
    R\text{-node}. 
\end{cases}
\end{equation}
Note that for each $x \in V(G)$
\begin{quote}
$w(x) = 1$ if and only if from $x$ there exists a winning
$R$-strategy. 
\end{quote}
Similarly, we define another Boolean-valued function $w' \co V(G) \to
\{ 0,1 \}$ such that
\begin{quote}
$w'(x) = 1$ if and only if from $x$ there exists a non-losing
$B$-strategy.  
\end{quote}
The definition of $w'$ is obtained from the definition of $w$ by
swapping $0$ with $1$ and $\lc{\and}$ with $\lc{\or}$.
Then, the theorem is equivalent to proving for $a$ -- the root of
$G$ -- that
\[
w(a) = \neg w'(a).
\]
It is actually easier to prove a stronger version of this, namely that
$w(x) = \neg w'(x)$ for \emph{all} $x \in V(G)$.
We do this by strong induction on the ``height'' $h_x$ of $x$ which is
defined by 
\[
h_x =
\begin{cases}
  0, & x \text{ leaf} \\
  1+\mathrm{max} \{ h_y \mid xy \in E(G) \}, & \text{otherwise.} 
\end{cases}
\]
\begin{itemize}[leftmargin=1.5em]

\item  
When $h_x = 0$, by the third hypothesis in the statement of the
theorem we have that $w(x) = \neg w'(x)$.

\item 
For the induction step, we assume $h_x = k > 0$.
We distinguish two cases: when $x$ is a $B$-node or when it is an
$R$-node.
\begin{itemize}[leftmargin=1.5em]

\item $x$ is $B$-node.
  Then
\begin{proofsteps}[justformat=\footnotesize\raggedleft]
\step*{}{$w(x) = \displaystyle{\bigwedge_{xy \in E(G)}}
  w(y)$}{definition of $w$}
\step*{}{\hspace{2.4em}$ = \displaystyle{\bigwedge_{xy \in E(G)}} \neg \
  w'(y)$}{\hspace{-5em}by the induction hypothesis since $h_y < k$}
\step*{}{\hspace{2.4em}$ = \neg \displaystyle{\bigvee_{xy \in E(G)}}
  w'(y)$}{DeMorgan laws}
\step*{}{\hspace{2.4em}$ = \neg w'(x)$}{definition of $w'$.}
\end{proofsteps}

\item When $x$ is $R$-node, the proof is similar to the above, just
  swap $\lc{\and}$ with $\lc{\or}$. 

\end{itemize}
\end{itemize}
\end{proof}

\section{Computing brute strategies}
\label{imp-sec}

The aim of this section is to specify the functions $w$ and $w'$ from
the proof of Theorem \ref{strategy-thm1} such that by instantiation we
obtain programs that can be executed for establishing the existence
of brute strategies in concrete situations.
The specification logic employed is (conditional) \emph{equational
  logic} with \emph{many-sorted algebra} (abbreviated \MSA)
semantics. 
In order to properly grasp the specifications/programs it is helpful
to have an understanding of the logical side.
For this reason, we will start this section with a very brief
presentation of equational logic with \MSA\ semantics.

\subsection{A quick reminder of equational logic in \MSA}

\MSA\ and its equational logic represent the traditional logic and
model theory of algebraic specification.
Any algebraic specification language, including those from the
OBJ-family, are built around \MSA, often extending it to more complex
logics and model theories.
Here we just review some basic definitions that are useful for getting
a very basic understanding of the math behind our specifications and
programs.  
For a deeper understanding, there are plenty of resources available,
such a \cite{sannella-tarlecki-book}.

\paragraph{\MSA\ signatures.}
We let $S^*$  denote the set of all finite sequences of
elements from $S$, with $[]$ the empty sequence. 
A(n $S$-\emph{sorted}) \emph{signature} $(S,F)$ is an
$S^* \times S$-indexed family of sets
$F = \{ F_{w \to s} \mid w \in S^*,\ s \in S \}$
of \emph{operation  symbols}.
Call $\sigma \in F_{[]\to s}$ (sometimes denoted simply $F_{\to s}$)
a \emph{constant symbol} of sort $s$.
A signature morphism $\varphi \co (S,F) \to (S',F')$ consists of a
function $S \to S'$ on the sort symbols and for each arity $w$ and
sort $s$ a function $F_{w\to s} \to F'_{\varphi(w) \to \varphi(s)}$. 

\paragraph{Equations.}
An \emph{$(S,F)$-term} $t$ of sort $s\in S$, is a structure of
the form $\sigma(t_1,\dots,t_n)$, where $\sigma \in F_{w\to s}$ and 
$t_1, \dots, t_n$ are $(S,F)$-terms of sorts $s_1 \dots s_n$, where
$w = s_1 \dots s_n$.
An \emph{(unconditional) ground $(S,F)$-equation} is an equality $t =
t'$ between $(S,F)$-terms $t$ and $t'$ of the same sort.
A \emph{conditional ground equation} is a sentence of the form
$\rho_1 \land \dots \land \rho_n \implies \rho$, where $\rho_1, \dots,
\rho_n, \rho$ are ground equations.
If $X$ is a finite set of variables for the signature $(S,F)$ then we
consider the extended signature $(S,F + X)$ that adjoins
the variables $X$ as new constants to $F$.
For any (potentially conditional) ground $(S,F + X)$-equation $\rho$,
$(\forall X)\rho$ is a \emph{universally quantified equation}. 

\paragraph{Algebras.}
Given a \emph{sort set} $S$, an $S$-\emph{indexed} (or \emph{sorted})
\emph{set} $A$ is a family $\{ A_s \}_{s \in S}$ of sets indexed by
the elements of $S$. 
Given an $S$-indexed set $A$ and $w = s_1...s_n \in S^*$, we let
$A_w = A_{s_1} \times \cdots \times A_{s_n}$; in particular, we let
$A_{[]} = \{\star\}$, some singleton set.
An $(S,F)$-\emph{algebra} (i.e., a model in \MSA) $A$ consists of 
\begin{itemize}

  \item an $S$-indexed set $A$ (the set $A_s$ is called the
    \emph{carrier} of $A$ of sort $s$), and

  \item a function $A_\sigma \co A_w \to A_s$ for each $\sigma \in
    F_{w\to s}$.

\end{itemize}   
If $\sigma \in F_{\to s}$ then $A_\sigma$ determines a point in $A_s$
which may also be denoted $A_\sigma$.
Any $(S,F)$-term $t = \sigma(t_1, \dots, t_n)$, where
$\sigma\in F_{w\to s}$ is an operation symbol and $t_1, \dots, t_n$
are $(S,F)$-(sub)terms corresponding to the arity $w$, gets
\emph{interpreted as an element} 
$A_t \in A_s$ in a $(S,F)$-algebra $A$ by 
$A_t = A_\sigma (A_{t_1}, \dots, A_{t_n})$. 

\paragraph{The satisfaction relation.}
The \emph{satisfaction relation} between  algebras and sentences,
denoted by $\models$, is the Tarskian satisfaction defined inductively
on the structure of sentences.    
Given a fixed arbitrary signature $(S,F)$ and an $(S,F)$-algebra $A$,  
\begin{itemize} 

\item $A \models t=t'$ if $A_t = A_{t'}$ for ground equations,

\item $A \models \rho_1 \land \rho_2$ if $A \models \rho_i$, $i=1,2$,
  and similarly for $\implies$, and 

\item for each $(S,F + X)$-equation $\rho$, 
    $A \models (\forall X) \rho$ if $A' \models \rho$ for each
    expansion $A'$ of $A$ with interpretations of the 
    variables of $X$ as elements of $A$. 

\end{itemize}

\paragraph{Initial semantics.}
A \emph{$(S,F)$-homomorphism} $h \co A \to B$, between $(S,F)$-algebras
$A$ and $B$, consists, for each sort $s\in S$, of a function $h_s \co
A_s \to B_s$, such that for each operation symbols $\sigma \in F_{s_1
  \dots s_n \to s}$ we have that
\[
  h_s (A_\sigma (a_1, \dots, a_n)) =
  B_\sigma (h_{s_1} (a_1), \dots, h_{s_n} (a_n)). 
\]
Given a class $\mathcal{C}$ of $(S,F)$-algebras, an initial algebra in
that class is any algebra $A \in \mathcal{C}$ such that for any $B\in
\mathcal{C}$ there exists an unique homomorphism $A \to B$.
A crucial result in algebraic specification is that for set $E$ of
equations (possibly conditional, possibly universally
quantified)\footnote{These are the sentences of traditional algebraic
  specification.} the class of the algebras satisfying $E$ has an
initial algebra.   
When $E$ is empty, this is the \emph{term algebra} which is
obtained by organising the set of the $(S,F)$-terms as an algebra in
the straightforward way.
When $E$ is not empty, the initial algebra is obtained as a
\emph{quotient} of the term algebra. 

\paragraph{Rewriting.}
This is the computational side of \MSA.
Given a set $E$ of (universally quantified conditional) equations with
some properties (in the literature called \emph{confluent} and
\emph{terminating} \emph{rewriting   systems}
\cite{terese,baader-nipkow1998}, etc.) we can compute \emph{normal
  forms} of terms. 
This can be considered a decision procedure for equality or as a form
of functional evaluation specific to equational logic.
The former is a computational equational logic perspective, while the
latter is a functional programming perspective on rewriting. 
Rewriting is crucial for our endeavour as it represents the execution
engine of our programs.
Rewriting is also intimately related to initial semantics \cite{bt-5}.
In other words, an equational specification that has the shape of a
rewriting system counts also as a program and can therefore be
executed to produce results. 

\subsection{The generic specification}

We can build an equational specification that computes the value
$w(a)$ from the proof of Theorem \ref{strategy-thm1} by recursion by
implementing formula \eqref{w-def1}.
Then, in the case of particular games we write specific equational
programs that compute the respective game tree.
These two parts are orthogonal to each other, which means they are
treated separately.
They inherently have different nature, as they belong to different
levels of abstraction.
This approach has `countless' benefits.
The abstract/generic part can be reused for any game.
Basically, for modelling computationally, or for programming, for any
other game one needs to address only the latter of the two parts and
plug it into the generic program.
There are also benefits regarding the clarity of the programming.
Moreover, in general, powerful modularisation techniques greatly
enhance the maintainability of the programs.

\paragraph{The data for programming generically formula
  \eqref{w-def1}.}
This consists of the following entities:
\begin{itemize}

\item The two players. 

\item A sort for the nodes in the game tree; since the nodes store
  configurations of the game this will be called \texttt{Config}.  

\item A Boolean-valued function that specifies the terminal
  configurations, i.e. the leaves of the game tree; this will be
  called \texttt{halted}.
  It is parameterised by the players in order to allow for partisan
  games. 

\item Another couple of Boolean-valued functions 
  that specify when $R$ ($\texttt{wonR}(p, c)$) and $B$
  ($\texttt{wonB}(p,c)$), respectively, can be declared a winner in a
  terminal configuration and the player $p$ was supposed to make the
  next move.     

\item A minimal specification for the sets of configurations; the sort
  of those will be called \texttt{SetConfig}.
  We do it only with a `union' operation \texttt{(\_,\_)} that has
  three of the four basic algebraic properties characterising set
  union, associativity, commutativity, and an identity element (the
  empty set called \texttt{empty}).
  The fourth property is idempotence, but we ignore that
  here.\footnote{So, strictly speaking, what we get are multi-sets
    rather than sets.}
  That each configuration is already a (singleton) set (of
  configurations) is specified by the \texttt{subsort} declaration.
  This allows for the building of the sets of configurations.
  The mathematics underlying the \texttt{subsort} declaration will get
  us beyond \MSA, it is an extension of \MSA\ called
  \emph{order-sorted algebra}.
  We will not do it here as, while the entire mathematical theory of
  \MSA\ can be lifted to the order-sorted algebra level (see
  \cite{osa-survey}), this leads to significant mathematical
  complications.  
  Moreover, here we will use this convenient technicality in a
  simple and rather intuitive way.    

\item Finally, we will have the function called \texttt{next-config},
  that actually builds the game tree by providing, for each
  configuration, the set of the configurations resulting from
  executing a game move.
  This is also parameterised by the players, to allow for partisan
  games.  

\end{itemize}

\begin{maude}
fth PLAYERS is
  sort Player .
  ops B R : -> Player .
  op opponent : Player -> Player .
  eq opponent(B) = R .
  eq opponent(R) = B .
endfth
\end{maude}

\begin{maude}
fth CONFIG is
  protecting BOOL .
  protecting PLAYERS . 
  sort Config .
  var p : Player .
  var c : Config .
  op halted : Player Config -> Bool .
  ops wonR wonB : Player Config -> Bool .
endfth
\end{maude}

\begin{maude}
fth SET-CONFIG is
  protecting CONFIG .
  sort SetConfig .
  subsort Config < SetConfig .
  op _,_ : SetConfig SetConfig -> SetConfig [assoc comm id: empty] .
endfth
\end{maude}

\begin{maude}
fth GAME-TREE is
  protecting SET-CONFIG . 
  op next-config : Player Config -> SetConfig .
endfth
\end{maude}




In the light of the \MSA\ definitions above, this \expmath\ code can
be understood rather easily, as the \expmath\ notations follow the
mathematical notations.
The attributes \texttt{assoc} and \texttt{comm} stand for the
associativity and commutativity equations, respectively, while
(\texttt{id : empty}) means that \texttt{empty}  is the identity for the
binary operation \texttt{\_,\_}.
The keyword \texttt{eq} designates an unconditional equation. 
Conditional equations are designated by \texttt{ceq}.

It is important to think of the models/algebras of these
specifications.
The keyword \texttt{fth} signals the \emph{loose} semantics, which
means that \emph{all} models are considered.
In spite of always considering only two players ($B$ and $R$),
\texttt{PLAYERS} is such a loose semantics module.
This is so because we want to have
$B$ and $R$ generic in order to be able to instantiate them. 
Not necessarily to give players other names, but mainly for being able
to switch the perspective of the game from one player to the other by
mapping $B$ to $R$ and viceversa
(and also some related operations such as \texttt{wonB} and
\texttt{wonR}) 
in the context of an appropriate `view'.\footnote{Later on we will see
  what this means.}
However, if we insist on models having exactly the players $B$ and $R$
then a solution is to write two more axioms:
\begin{maude}
  var P : Player .
  eq (P == B) or (P == R) = true .
  eq (B == R) = false .
\end{maude}
The models of \texttt{CONFIG} expand the standard model of the
Booleans and the models of \texttt{PLAYERS} (this is the meaning 
of \texttt{protecting}) with all possible interpretations of the other
entities of the signature of \texttt{CONFIG}.
These `loose' interpretations allow for the possibility to model
various different games.
In other words, at this stage, \texttt{Config}, \texttt{halted},
\texttt{wonR}, \texttt{wonB}, and \texttt{next-config} are
under-specified in the sense that they are left completely abstract.
The process of obtaining an executable program for any concrete game
consists mainly of making these entities concrete by appropriate
specifications.  

\paragraph{The coding of formula \eqref{w-def1}.}
The next module is the core module of our generic equational
specification.
In the case of the concrete games played by Benny and Rebecca, this
allows for computing the existence of winning strategies for Rebecca.
\begin{itemize}

\item   
The intended meaning of $w(p, c)$ is that it gives true if and only if
Rebecca has a winning strategy from the configuration $c$ when the
player $p$ has to move.

\item 
The operation \texttt{w-aux} specifies a computation by recursion of
the conjunctions and the disjunctions from formula \eqref{w-def1}. 
At this stage we cannot do any computation, but this will become
possible when we instantiate the generic module. 

\item
The module is \emph{parameterised} by the module \texttt{GAME-TREE},
meaning that in order to obtain programs for concrete games we have to
make the entities specified by \texttt{GAME-TREE} concrete.

\end{itemize}
Otherwise, in the light of formula \eqref{w-def1}, the module
\texttt{WINS} below is pretty straightforward.
Only \texttt{EXT-BOOL} and its entities \texttt{and-then} and
\texttt{or-else} needs additional explanations, which will be provided
below. 
\begin{maude}
fmod WINS{X :: GAME-TREE} is
  protecting EXT-BOOL .
  op w : X$Player X$Config -> Bool .
  op w-aux : X$Player X$SetConfig -> Bool .
  var p : X$Player . 
  var c : X$Config .
  var S : X$SetConfig .
  eq w(p, c) = if halted(p, c) then wonR(p, c)
                               else w-aux(opponent(p), next-config(p, c)) 
               fi .
  eq w-aux(R, empty) = true .
  eq w-aux(R, (c, S)) = w(R, c) and-then w-aux(R, S) .
  eq w-aux(B, empty) = false .
  eq w-aux(B, (c, S)) = w(B, c) or-else w-aux(B, S) .
endfm
\end{maude}
The keyword \texttt{fmod} denotes \emph{initial} semantics, which in
this case is explained by the concept of \emph{persistent adjunction}
\cite{ins,iimt2}.
Here we will not further discuss this mathematics, as it is beyond the
scope of this paper. 
The semantics of \texttt{WINS} is given by all models of
\texttt{GAME-TREE} expanded with interpretations of \texttt{w} and
\texttt{w-aux} as Boolean-valued functions. 
For the models of \texttt{GAME-TREE} that do correspond to finite
games, there are unique interpretations of these functions, these being
determined by the situations at the level of the leaves of the game
tree.
To achieve completeness at the level of the leaves, in the programs
for the concrete games, at each leaf \texttt{halted}, \texttt{wonR},
and \texttt{wonB}, have to be evaluated as either \texttt{true} or
\texttt{false} (semantically, this requirement is already specified by
the \texttt{protecting} imports). 
This is what the equations of \texttt{WINS} give us.

The use of \texttt{if\_then\_else\_fi} is essentially an abbreviation
for the corresponding set of two conditional equations. 
The Boolean connectives \texttt{and-then} and \texttt{or-else} are
used instead of the standard \texttt{and} and \texttt{or},
respectively.  
This is a helpful facility provided by \expmath\ that may speed up the
computation as follows.  
If the recursive computation of a $\texttt{w-aux}(R,\dots)$ encounters a
\texttt{false} value of a $\texttt{w}(R,\ldots)$ then the recursion
process exits immediately with the result \texttt{false}.   
This shortcut eliminates the pointless computations of other
$\texttt{w-aux}(R,\dots)$, thus speeding up the main evaluation of $w$.
This is taken care by \texttt{and-then} which has the same logical
semantics as the simple conjunction \texttt{and}, but is
computationally more efficient.
The operation \texttt{and-then}, together with its disjunctive
counterpart \texttt{or-else} (which gives a result immediately after
finding a \texttt{true} value), are available only if we
explicitly import the module \texttt{EXT-BOOL} (which explains the
second line of \texttt{WINS}).

Those readers who have background in game theory may recognise
\texttt{WINS} as a generic equational logic axiomatisation of the
methodology to analyse games belonging to certain classes, known by
the name of \emph{backward induction} \cite{fudenberg-tirole91}, which
is traditionally performed by hand. 

\section{Programs for concrete games}
\label{concrete-sec}

In this section we show how the generic specification \texttt{WINS}
can be instantiated to obtain equational programs for concrete games.
We will illustrate this in some detail with the game of Example
\ref{heap-game}, and in the cases of the other examples introduced in
Section \ref{gamesex} we only present the main ideas of the respective
instantiations. 
In the final part of this section we discuss methods to fight the
complexity of the game trees.  

\subsection{The heaps game}

\paragraph{Interpreting the parameter.}
In the context of the heaps game, we have to provide the definitions
for \texttt{Config}, \texttt{halted}, \texttt{won}, and
\texttt{next-config}. 
Since the programs for these are of secondary interest, we will not
present them now, but rather exile them to the appendix.
However, we review their main ideas.
\begin{itemize}

\item The interpretation of \texttt{Config} consists of pairs of
  heaps, heaps being represented by their sizes.
  This data type is called \texttt{TwoHeaps}. 

\item A configuration is terminal if and only if it is $(0,0)$. 

\item $\texttt{wonR}(p, c)$ is true if and only if $p = B$ and
  $c$ is terminal.

\item The definition of \texttt{next-config} (called
  \texttt{next-heaps}) implements the actual moves of the game
  according to their specification from Example \ref{heap-game}.
  This involves a couple of auxiliary operations. 
  
\end{itemize}

\paragraph{Instantiating the generic program.}
Now, that we have the concrete instances for the abstract entities of
the parameter \texttt{GAME-TREE}, we can proceed with its instantiation
by the following \emph{view}.\footnote{\emph{View} is standard
  terminology in algebraic specification for instances of module
  parameters.
  This will be explained during what follows, as part of the
  explanation of the parameter institution mechanism.}
Note that we do not need to write anything for \texttt{halted},
\texttt{wonR} and \texttt{wonB}, because their instances have exactly
the same name, and because of that, \expmath\ does the respective
mappings by default. 
Note that for this particular game the set of the `next'
configurations (i.e., the successor nodes in the game tree) does not
depend on which player has to do the move.
This is reflected in the definition of \texttt{next-config}, the
argument $p$ being `dead'. 
\begin{maude}
view HeapsGame from GAME-TREE to HEAPS-GAME is
  sort Config to TwoHeaps .
  sort SetConfig to Set{TwoHeaps} .
  op next-config(p:Player, c:Config) to term next-heaps(c:TwoHeaps) .
endv
\end{maude}
The instance of \texttt{WINS} for our concrete game is thus obtained: 
\begin{maude}
fmod WINS-HEAPS is
  protecting WINS{HeapsGame} .
endfm
\end{maude}
The specification \texttt{HEAPS-GAME} is presented in Appendix
\ref{heaps-appendix}. 

\paragraph{The pushout.}
This instantiation can be visualised by the following diagram:
\[
  \xymatrix @+3ex {
  \texttt{GAME-TREE} \ar[r]^{\hspace{-2em}\subseteq} \ar[d]_{\texttt{HeapsGame}} &
  \texttt{WINS\{X :: GAME-TREE\}} \ar@{.>}[d] \\
  \texttt{HEAPS-GAME} \ar@{.>}[r]_\subseteq & \texttt{WINS\{HeapsGame\}}
  }
\]
The proper reading of this diagram takes us to the mathematical
foundations of parameterised programming in the OBJ-family of
languages, \expmath\ included.
These foundations are based on category theory \cite{maclane98} and
institution theory
\cite{ins,iimt2,sannella-tarlecki-book,sannella-tarlecki88,AlgStrucSpec,modalg}.  
The diagram represents a `pushout' in the category of \MSA\
specifications/programs.
All arrows represent \emph{specification morphisms}, which are
generalised signature morphisms\footnote{Signature morphisms map sorts
  to sorts and operation symbols to operation symbols in a coherent
  way.
  Views generalise the signature morphisms by allowing operation
  symbols to be mapped to terms, such that the arities match through
  the mapping of the sorts.
  The mapping of the operation symbols may `forget' arguments (like
  \texttt{HeapsGame} does) but is not allowed to bring in new
  arguments.}
such that all models of the target specification
represent also models of the source specification when they are
`reduced' to models of the source signature. 
The upper arrow, labelled by an inclusion, represents the fact that
the parameter \texttt{GAME-TREE} is a part of the abstract program
\texttt{WINS}.
The left arrow represents the fact that the `view' \texttt{HeapsGame}
is a proper mapping of the entities of the parameter
\texttt{GAME-TREE} to corresponding entities of the module
\texttt{HEAPS-GAME}. 

\paragraph{The semantics of the instantiation.}
We are interested in the semantics of \texttt{WINS\{HeapsGame\}}.
This is based on the concept of model amalgamation from algebraic
specification theory.
It goes like this.
Any model of \texttt{WINS\{HeapsGame\}} comes as an `amalgamation' of a
model of \texttt{HEAPS-GAME} and a model of \texttt{GAME-TREE}.
These two models should be mutually coherent, meaning that they share
their parts corresponding to \texttt{GAME-TREE}.
Since \texttt{HEAPS-GAME} is an initial semantics module, it has only
one model (up to isomorphism), which essentially is the game tree.
This gives us that \texttt{WINS\{HeapsGame\}} also has one model
which is the same game tree but now enhanced with the Boolean-valued 
functions \texttt{w} and \texttt{w-aux}, which, as we discussed above,
have unique interpretations. 
Of course, all these have rigorous mathematically detailed
explanations within the theory of algebraic specification, such as in
\cite{sannella-tarlecki-book,modalg,AlgStrucSpec}.

The full code of this example is available in the repository
\cite{heaps-game}. 

\subsection{The board game}
In all cases, including the board game, the instantiation mechanism is
the same with that described for the heaps game, by the pushout technique. 
A real difference between these two instantiations occurs when
we look at their complexity and sophistication; in the case of the
board game both complexity and sophistication are much greater than
for the heaps game.  
Unfortunately, this situation impacts negatively the efficiency of the
computations.
For instance, in the case of the board game, the specification of the
instantiation view makes a rather intensive use of the predefined data
type of sets, which requires rewriting \emph{modulo} associativity and
commutativity.
Then 
\begin{itemize}

\item The instantiation of \texttt{Config} uses a representation of
  the board as sets of cells (positions), which are represented by
  their coordinates on the board.
  Then, a configuration of the game is represented as a triple
  $(P, P_2, k)$ where $P$ represents the set of the free positions on
  the board, $P_2$ the set of the squares $2 \times 2$ in $P$, and $k$
  is the difference between the blue and the red cells that have already
  been coloured.
  While $P$ carry all necessary information, meaning that $P_2$ and
  $k$ can be computed from $P$, we chose to include $P_2$ and $k$
  explicit and keep them updated in the configurations in order to use
  them directly without having to compute them in duplicates.

\end{itemize}
With such interpretation of \texttt{Config} we get the following
further instantiations for the operations of \texttt{GAME-TREE}: 
\begin{itemize}

\item $\texttt{halted}(p, (P, P_2, k)) = (|P| \leq k) \lor \big( (p=
  B) \land (P_2 = \emptyset) \big)$.
  The former case means that there are not enough free positions left
  to bridge the gap between $B$ and $R$, so that $B$ wins.
  The latter case means that $B$ has to move but cannot.

\item $\texttt{wonR}(p, (P, P_2, k)) = (k < |P|) \land
  \texttt{halted}(p, (P, P_2, k))$.
  So, if the game halts and the gap between the blue and the red cells
  is less than the free cells, then $R$ wins. 

\item $\texttt{wonB}(x) = (\neg \texttt{wonR}(x)) \land
  \texttt{halted}(x)$. 

\item The instantiation of \texttt{next-config} is obtained by a
  recursive computation based on the specific moves of the board game,
  that is supported by an auxiliary function in the style of how
  $\texttt{w-aux}(\_)$ supports $\texttt{w}(\_)$ in \texttt{WINS}. 
  
\end{itemize}
The actual Maude code of this instantiation is available in the
repository \cite{board-game},
where all missing details can be studied.

\subsection{The transition systems game}

The appropriate way to specify the bisimulation problem is in two
abstraction stages by considering also the transition systems as
parameters.
This double-parameterised module can be thus used as an instance of
the parameter of \texttt{WINS}.
The full code of the bisimulation game instantiation, together 
with some runs that decide some bisimilarities, can be found in
\cite{bisimilarity-game}. 
The main ideas of this instantiation are as follows.
\begin{itemize}

\item We use distinct parameters for the two transition systems in
  order to allow their states to belong to possibly different
  domains. 

\item We specify transition systems in the fashion of a succesor
  function:
  \begin{maude}
  op succ : State TS -> StateSet .
  \end{maude}
  where \texttt{State} is the sort of the states ($S$), \texttt{TS} is
  the sort of the transition system, and $\texttt{succ}(s, \mathit{TS})$
  specifies the set $\{ s_1 \mid (s, s_1) \in T \}$ (where $\mathit{TS} =
  (S, T)$), i.e. all the states $s_1$ such that there exists a
  transition $\xymatrix{s \rto & s_1}$. 

\item Configurations are tuples consisting of two transition systems
  and two sequences of transitions represented by the respective
  sequences of states (one for each transition system).

\item The operations \texttt{halted}, \texttt{wonB}, \texttt{wonR},
  \texttt{new-config} are specified in modules parameterised by the
  two abstract transition systems.

\item The instantiations of \texttt{WINS} with concrete bisimulation
  games are obtained in two stages:
  \begin{enumerate}

  \item By the standard pushout technique, we instantiate
    \texttt{GAME-TREE} with the module \\
    \texttt{NEXT-CONFIG\{T1,T2\}} which has two transition systems as
    parameters and which is an abstract specification of the
    bisimulation game trees. 
    \[
      \xymatrix @C+7ex @R+3ex {
      \texttt{GAME-TREE} \ar[d]|{\texttt{TS\{T1 :: TRANSITION-SYSTEM,}
      \texttt{T2 :: TRANSITION-SYSTEM\}}} \ar[r] &
      \texttt{WINS\{X :: GAME-TREE\}} \ar@{.>}[d] \\
      \texttt{NEXT-CONFIG\{T1,T2\}} \ar@{.>}[r]_\nu &
      \texttt{WINS\{TS\{T1,T2\}\}}
      }
    \]
    The view \texttt{TS} maps the sort \texttt{Config} to a sort of
    configurations that are tuples consisting of two transition
    systems and two sequences of states (one for each transition
    system).
    Details of these are in \cite{bisimilarity-game}.

  \item Then, we consider two specifications of concrete transition
    systems, \texttt{CTS1} and \texttt{CTS2}, with the corresponding
    views from \texttt{TRANSITION-SYSTEM}.
    Based on these views, we instantiate the parameters \texttt{T1}
    and \texttt{T2} by the following colimit:
    \[
      \xymatrix{
        \texttt{TRANSITION-SYSTEM} \ar[r] \ar[d] &
        \texttt{NEXT-CONFIG\{T1,T2\}} \ar[d]^\nu  &
        \texttt{TRANSITION-SYSTEM} \ar[d] \ar[l] \\
        \texttt{CTS1} \ar@{.>}[dr] &
        \texttt{WINS\{TS\{T1,T2\}\}} \ar@{.>}[d] &
        \texttt{CTS2}  \ar@{.>}[ld] \\
        & \texttt{WINS\{TS\{CTS1,CTS2\}\}} & 
      }
    \]
    In \cite{AlgStrucSpec} it is explained how the above colimit can
    be computed as two pushout squares.

  \end{enumerate}
  
\end{itemize}

\subsection{Fighting the complexity of computations}

The heaps game, on the one hand, and the board and the transition
system games, on the other hand, differ drastically in the aspect of
complexity of computation. 
In the former case computations run fast, while in the latter cases
the complexity of the computations explode. 
This contrast is emblematic for our method.
We can distinguish two factors that cause the issue with complexity.
\begin{itemize}

\item A first factor comes with a big size of the game tree, which
  is determined by a big number of possibilities to perform a move.
  In other words, a big $\texttt{next-config}(p, c)$ leads to complexity
  of computations issues.
  In both the heaps and the board games the dynamics of
  $\texttt{next-config}$ is similar in the sense that it decreases in
  size as we advance in the game.
  But this is not necessarily true for all combinatorial games.
  The problem with the board games (but there are also other classes
  of combinatorial games that behave similarly) is that we often get a
  lot of possibilities for the first few moves, that cumulate quickly
  and lead to huge computation volumes.
  This is exactly what happens with our board game, and which does not
  happen with the heaps game.

\item Another factor comes with the cost of the computations, which
  can be expensive. 
  This is the case of our board game, but not of the heaps game.
  In the latter case a move means a simple numerical subtraction,
  while in the former case one has to subtract some positions, or even
  sets of positions from the board (or, more precisely, from the
  current set of the free positions on the board).
  Operations on sets are much more expensive than operations on
  integers. 

\end{itemize}
The worst scenario is when both above mentioned factors are not
favorable, which is the case with our board game.
The best scenario is the opposite one; this applies to the heaps game.
In between there are two mixed scenarios.
To get an idea about the differences, in the case of the heaps game,
by 5M rewrites in a third of second we get all the winning
configurations $(k,n)$, $k,n \leq 100$ for the second player, while in
the case of the board game by 2M rewrites in less than a third of a
second we get the answer for the game on the $5 \times 5$ board.
But then the complexity explodes, for the $6 \times 6$ board it takes
1.6B rewrites in three minutes to get the result, and the $7 \times 7$
board is hopeless.

Below is a(n incomplete) list of techniques that may help to speed up
the executions of the programs that implement our method in specific
cases.
\begin{itemize}

\item Exploiting the symmetries of the game tree may reduce it to a
  fraction of itself. 
  An example is given by our board game, in which, due to the three
  symmetries of the board (vertical, horizontal, and one diagonal),
  the first move may be considered in only one half of one of the four
  quarters of the board. 
  Unfortunately, this does not iterate to the further moves down this
  game.
  In the situation when symmetries do occur in all states of a game,
  the effect of exploiting them in the programming can be dramatic.
  The heaps game also has an aspect of symmetry, which is different
  than the previously discussed kind of symmetry, namely that the
  heaps can be considered ordered by their size.
  However, this does not have the same effect on the game tree like
  the board symmetries. 
  The reader can check this on the game tree of Figure
  \ref{a812-1fig}.  

\item Maude allows for user-defined data types through initial
  semantics and provides great computational support for this,
  including rewriting modulo axioms such as associativity,
  commutativity, etc.
  However, such computations may be quite expensive.
  For instance, computations with sets are more expensive than with
  lists.
  This means that the benefits of a highly declarative specification
  style should be balanced with a choice of more
  computational-friendly data types.
  A good development may do both in parallel. 

\item A skilled programming style helps to fight complexity.
  For instance, we may avoid computing duplicates by defining
  configurations that carry redundant data, such as in the case of the
  board game where $P_2$ and $k$ could be computed from the $P$. 
  
\item Smart data representation may also help to reduce the complexity
  of computations.
  For instance, in the case of our board game we represented the
  squares $2 \times 2$ by the upper-left corner. 
  
\end{itemize}

There is also another kind of unfavourable aspect that comes with our  
proposal.
Our method is logic-based, algebraic, and highly declarative.
On the one hand, this means clarity, rigour, solid foundations.
On the other hand, sometimes this may mean a certain sacrifice in
efficiency.
Everything can be of course done with some fast imperative programming
language, but by sacrificing precisely the good features mentioned
above.
In general, it is difficult to balance the gains and loses between the
two choices, but we believe that the inherent computational complexity
of a certain game cannot be addressed in a substantial way by
switching between programming paradigms.
From this perspective it is good to stick with logic-based solid
foundations, with declarative programming.
Having said that, there is always the possibility to mix them in the
sense of having a prototype based on our method, which can be further
implemented in a faster programming environment.
In the formal specification jargon this is called \emph{refinement}.
A middle ground for refinement is to ``refine'' the `pure'
logic-based programs/specifications to programs/specifications that
are still declarative but employ some imperative-styled features, such
as the \emph{matching equations}. 

Finally, it is important to think pragmatically about fighting
complexity.
If something serves the purpose without a special effort to reduce the
complexity of the executions, then why bother.
For instance, in the heaps example a straightforward program is enough
to compute fast a lot of data to use for experimental purposes. 

\section{Formulating and proving smart strategies based on
  experiments}

By design, the scope of \texttt{WINS} is to establish the existence of
winning strategies for particular concrete games.
In many cases, that is enough.
For instance, in the bisimulation game, in most cases we are just
interested to establish bisimilarity for particular concrete
transition systems, which means (the respective instance of)
\texttt{WINS} alone serves the purpose. 
But, in experimental contexts, \texttt{WINS} can also play a great
role for ``conjecturing'' smart winning strategies. 
In this section we will discuss three techniques for this, illustrated
on our heaps and board games, as follows:
\begin{enumerate}

\item Sometimes, by running an instance of \texttt{WINS} we can get
  relatively fast a lot of experimental data that gives good
  information about the winning positions.
  By analysing such data we can understand winning strategies,
  formulate them as conjectures, and prove them mathematically.
  This is the case with our heaps game. 

\item The board game has the opposite characteristics of the heaps
  game, its complexity explodes quickly and, moreover, what we get
  from direct experimentation with the corresponding instance of
  \texttt{WINS} does not provide the same level of information about
  winning strategies like in the case of the heaps game.
  However, based on what we get, we can run through winning strategies
  and detect patterns for the winning moves.
  This techniques relies on non-deterministic rewriting, which is a
  general programming technique in which rewriting is used with a
  different meaning than in equational programming.
  We will provide a brief introduction to its semantics. 

\item Alternatively to the previous technique, we can obtain traces of
  winning strategies by equational programming, by upgrading the
  generic module \texttt{WINS} from Boolean-valued results to such
  traces.
  We will present such a generic upgrade that can be used in the same
  way we use \texttt{WINS}, and where traces are also treated
  abstractly, allowing for manifold interpretations. 

\end{enumerate}
Each of these three techniques have a general character, meaning that
they can be applied to other games than those discussed here.
Given an actual game, everything depends on understanding which is the
most appropriate technique to be applied. 

\subsection{The simple application of \texttt{WINS}}

\begin{figure}
\begin{center}
\scalebox{1}{
\includegraphics{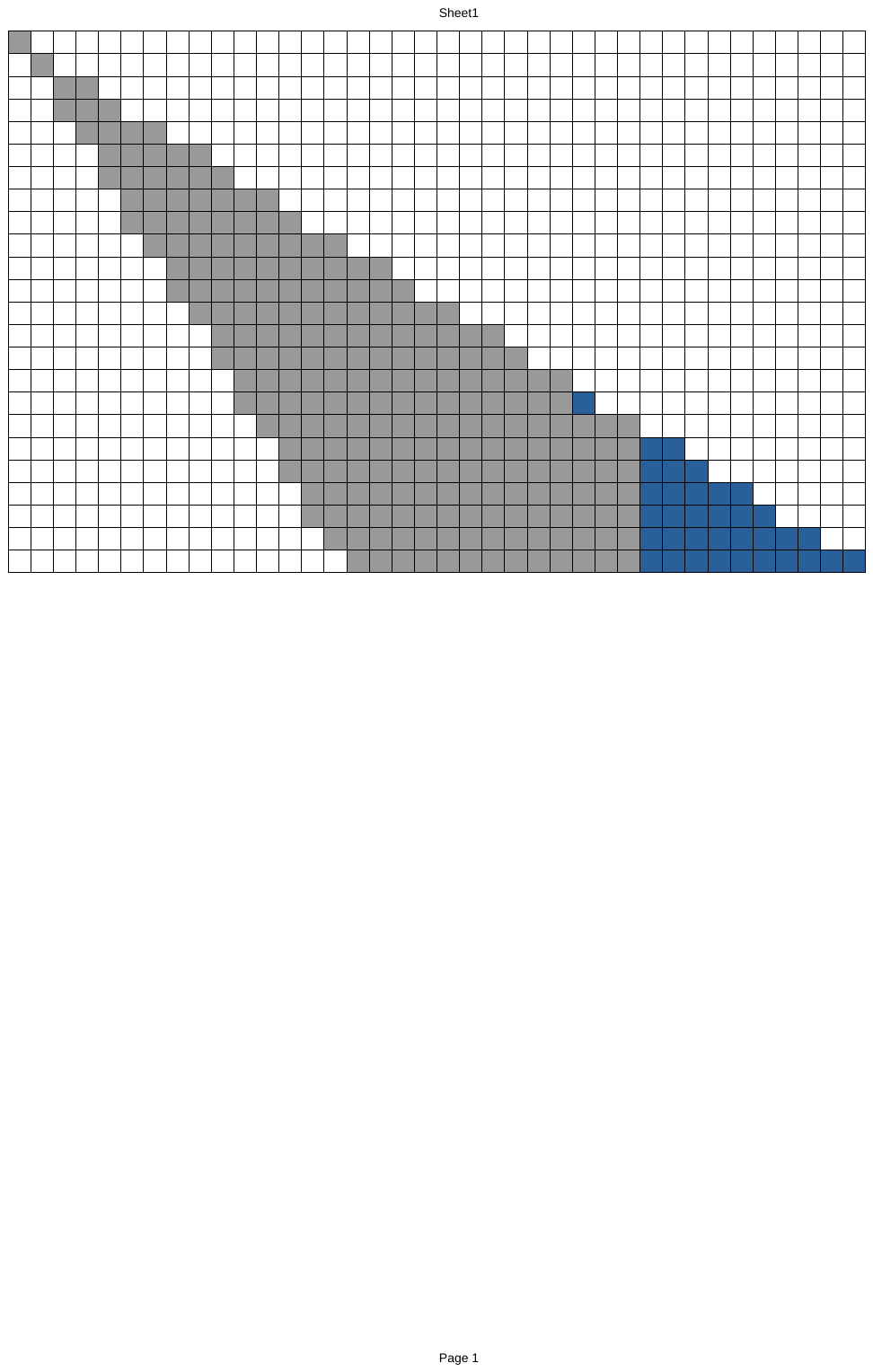}
}
\caption{Winning/losing positions for the heaps game.}
\label{a812-positions}
\end{center}
\end{figure}

By using \texttt{WINS\{HeapsGame\}} we are able to compute very fast 
$w(k, n)$ for $1 \leq k, n \leq 100$.
We wrote some short auxiliary programs in order to do this fully
automatically. 
An initial segment of the result may be visualised in the following
two-dimensional grid shown in Figure \ref{a812-positions}, where the
grey cells represent the pairs of heaps for which Benny has a winning
strategy. 

The inspection of the result showed in Figure \ref{a812-positions}
reveals the following facts:
\begin{itemize}

\item The winning configurations for Benny are grouped together, so
  for each $n$ there exists $a_n, b_n$ such that $(k,n)$ is a winning
  configuration for Benny if and only if
  $a_n \leq k \leq b_n$.

\item $a_n \leq n \leq b_n$.

\item $b_n - a_n = n - 1$. 
  
\end{itemize}
At this stage we have to determine some formulas for $a_n$ and $b_n$
and then formulate explicitly a strategy and prove its validity.
The former task may be quite difficult, but we have a powerful tool
called \emph{The Online Encyclopedia of Integer Sequences} \cite{oeis}.
From the results of our computing experiments we obtain that the
initial segments of $(a_n)_{n\in\omega}$ and of $(b_n)_{n\in \omega}$
are 
\begin{itemize}

\item[]
  $0,1,2,2,3,4,4,5,5,6,7,7,8,9,9,10,10,11,12,12,13,13,14,15,\ldots$,
  and 

\item[]
  $0,1,3,4,6,8,9,11,12,14,16,17,19,21,22,24,25,27,29,30,32,33,35,37,\ldots$,
  respectively.  

\end{itemize}
The database of this encyclopedia suggests that
$a_n = \ceiling{n/\phi}$ and $b_n = \floor{\phi n}$ where
$\phi = \frac{1+\sqrt{5}}{2}$ is the famous ``golden ratio''.
Recall that $\phi$ is the positive root of $x^2 - x -1$, the other
root being $-1/\phi$. 
Thus, the set of the winning positions for Benny, when Rebecca makes
the first move, seems to be
\begin{equation}\label{l-def}
  \{ (k,n) \in \omega^2 \mid n/\phi \leq k \leq \phi n \}
\end{equation}
(by $\omega$ we denote the set of the natural numbers). 
It remains to prove this. 
    
\paragraph{An smart strategy.}
In  Figure \ref{a812-positions} let $W$ denote the white zone and $L$
the grey zone.
Thus
\[
  L = \{ (k,n) \in \omega^2 \mid n/\phi \leq k \leq \phi n \}
  \mbox{ \ and \ }
  W = \omega^2 \setminus L = \{ (k,n) \in \omega^2 \mid \phi k <
  n \text{ or } \phi n < k \}.
\]
Then, $W$ should represent the winning positions for the player who is
about to move, while $L$ represents the set of the losing positions.
That holds if the following things happen:
\begin{enumerate}[leftmargin=1.5em]

\item $(0,0) \in L$. 

\item From $W$ a player \emph{can} always move to $L$. 

\item From $L$ it is possible to move \emph{only} to $W$. 
  
\end{enumerate}
We can have a mathematical proof of these as follows.
First, note that in all inequalities involving $\phi$, unless $(k,n) =
(0,0)$, because $\phi$ is irrational we can interchange $\leq$ and
$<$.
We will also use the equivalence between $n/\phi < k < n\phi$ and
$k/\phi < n < k\phi$. 

\begin{prop}
1., 2., 3. hold true.
\end{prop}

\begin{proof}
By symmetry we may assume that $k \leq n$.
\begin{enumerate}[leftmargin=1.5em]

\item This is obvious. 

\item Let $(k,n) \in W$.
  Since $k \leq n$, it follows that $\phi k < n$ because $(k,n) \in W$
  means that $\phi k < n$ or $\phi n < k$. 
  Since $\phi k - k/\phi = k\frac{\phi^2 -1}{\phi} = k$, it follows
  that there exists $k/\phi < r < \phi k$ such that $n \equiv r \mod
  k$.
  Since $n > \phi k$, we have that $n \not= r$.
  Hence the player can move to $(r,k)\in L$. 

\item Let $(k,n) \in L$.
  We may assume that $(k,n) \not= (0,0)$, otherwise the game is
  already done.
  Since $k < n < \phi k < 2k$, the only possible moves are to $(0,n)$
  and to $(k, n-k)$.
  Because $n > 0$,
  $(0,n) \in W$.
  Hence, it remains to prove that $(k, n-k) \in W$ too.
  Indeed, $n < \phi k$ implies $n < (1 + 1/\phi)k$ (since
  $\phi = 1 + 1/\phi$) which further implies $n-k < k/\phi$, hence
  $n-k \not\in (k/\phi, \phi k)$. 
  
\end{enumerate}
\end{proof}

Now, to formulate a smart strategy is straightforward. 
Indeed, since $(0,0) \in L$, it follows that any player who has to
move from $W$ wins just by always moving to $L$.
This is what Benny has to do as the second player since after the
first move by Rebecca, if she starts from $L$, then he will be in
$W$.
From $L$, Rebecca loses and Benny wins if he pursues this strategy.
Conversely, if $(k, n) \in W$, i.e. starts from $W$, then she wins and 
Benny loses.

\begin{cor}
Benny has winning strategies when the heaps have $k$ and $n$ tokens,
respectively, such that $n/\phi < k < \phi n$.
\end{cor}
Finally, we can now revisit the example of Figure \ref{a812-1fig} and
note that $(2,5) \in W$, indeed.
This means that Rebecca, the first to move, has a winning strategy.
We note also that the only winning strategy for Rebecca, as outlined
after Definition \ref{strategy-dfn1}, corresponds to the general $W-L$
formulation for winning strategies.
This situation is emblematic for impartial games. 

\subsection{By non-deterministic rewriting, running through winning
  strategies }
\label{poa-sec}

Let us consider our board game.
By running the appropriate instance of \texttt{WINS} we get that, up
to the $6 \times 6$ board, Rebecca has winning strategies if and only
if the size of the board is an odd number.
We cannot go beyond $6$ with the current computing power limitation of
our machines as the complexity of the game explodes badly.
As we discussed in Section \ref{concrete-sec}, in the case of this
game we can exploit the board symmetries and divide the complexity by
a factor of eight, but we can do it only once, with the first move of
the game.
However, even these five cases are enough to have a high degree of
confidence in the validity of the conjecture that $R$ has winning
strategies if and only if the size of the board is an odd number. 
But this information is not enough to indicate some smart winning
strategy, that would also support a proof of the conjecture.
What we can do is to run the game but only through the winning
configurations/positions, and analyse patterns for the moves leading to
such configurations.
Then, eventually, we may understand the winning strategies, and that
understanding would lead to completely ``solving'' the game. 

\paragraph{What is non-deterministic rewriting?}
This method requires \emph{non-deterministic rewriting}, a programming
paradigm beyond equational logic but which shares the same basic
algebraic rewriting algorithm.
Its main characteristics are:
\begin{itemize}

\item Rewriting is not used to compute values in data types, like in
  equational programming, but rather to compute \emph{states} of
  algebraic transition systems.
  This means that the requirement of the confluence of the rewriting
  systems, and even termination is dropped off.\footnote{This does not
    mean that these properties are not allowed; they remain desirable
    but to a much lesser extent.}
  This gives rewriting a non-deterministic aspect.

\item Its logic is \emph{preordered algebra} (abbr. $\POA$), which is
  an extension of $\MSA$ both at the syntactic and at the semantic
  levels as follows:
  \begin{itemize}

  \item It shares the concept of signature with $\MSA$, but introduces
    a new type of atomic sentences, the \emph{transitions}, which are
    of the form $t \trans t'$, with $t$ and $t'$ being terms of the
    same sort. 

  \item The $\POA$-models are $\MSA$ algebras $A$ enhanced, for each
    sort $s$ of the respective signature, with a preorder relation
    $\leq_s$ on $A_s$, which is preserved by the operations of $A$.

  \item $A \models t \trans t'$ if and only if $A_t \leq A_{t'}$. 
    
  \end{itemize}

\end{itemize}
In Maude, the (universally quantified conditional) transitions are
called \emph{rules} (keywords (\texttt{c})\texttt{rl}), while in
CafeOBJ are called just \emph{transitions}.
The common semantics of Maude \cite{maude-book} (and also the original 
semantics of CafeOBJ \cite{caferep}) are based on a generalisation of
$\POA$ that considers categories rather than preorders.
However, the upgraded semantics of CafeOBJ (reported in
\cite{cafefun}) is based on $\POA$.\footnote{One of the main reasons of
  that upgrade was alignement with institution theory
  \cite{ins,sannella-tarlecki-book,iimt2}, etc., and its approach
  to modularisation.}
There are some significant differences between the two semantics and
also about how Maude and CafeOBJ realise non-deterministic rewriting.
With the exception of some aspects of the module system, these
differences do not have any meaning for our work, so we will not
further discuss them.
Therefore, we will stick with the simplest semantics of transitions,
as provided by $\POA$. 

\paragraph{Running two transitions.}
The moves of the game can be specified by a transition for each of the
two players.
Below we show how this can be done generically, by leaving some data
underspecified.
The following two transitions are written for running the situation
when $R$ moves on winning configurations only.
For $B$ something similar can be written.
\begin{maude}
crl < B | c | lB | lR > => < R | cB | lB+ | lR  > 
    if not halted(B, c) . 
crl < R | c | lB | lR > => < B | cR | lB  | lR+ > 
    if (not halted(R, c)) and w(B, cR) .
\end{maude}
Here \texttt{c : Config} while \texttt{lB, lB+, lR, lR+} are the
currently accumulated lists of moves of $B$ and $R$.  
A move by $B$ upgrades \texttt{lB} to \texttt{lB+}, and
similarly for $R$.
How this works for our board game can be seen in the code in the 
repository \cite{board-game}.
Depending on the optimal kind of information  corresponding to the
actual game, \texttt{lB, lB+, lR, lR+} can be another data type rather
than lists, such as sets.  

\begin{figure}[H]
\begin{center}
\scalebox{.7}{
\includegraphics{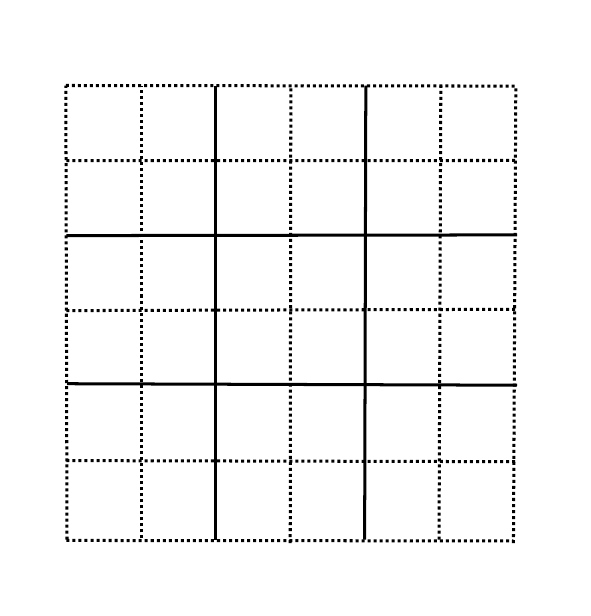}
}
\caption{$3 \times 3$ grid on the $6 \times 6$ board} 
\label{grid2x2}
\end{center}
\end{figure}

In the case of our board game, we already know that $R$ has
winning strategies on the $5 \times 5$ board.
By running such code for $R$ on the $5 \times 5$ board we 
do not get understandable data.
Then, we can turn to the other player, $B$.
We know that he has ``winning'' strategies (which in his case means
just that he does not lose) for the even sizes of the board, such as
$6 \times 6$. 
So, we run the respective two transitions on the $6 \times 6$ board,
for all possible first two moves of $B$ (for more moves the complexity
would be already too much),\footnote{
  For the three possibilities for the first move the computation
  takes 5.3B rewrites (in ten minutes) and for the other 828
  possibilities for the first two $B$-moves it takes another 50B
  rewrites (in 100 minutes).
}
also by dividing the space of the states by eight according to the
above discussed three symmetries.  
By analysing the total of 831 possibilities, we could notice the
following pattern: the first move was \emph{always} on the grid shown
in Figure \ref{grid2x2}, while in $83\%$ of the cases the second move
was also on the same grid.
Moreover, all of the remaining $17\%$ cases are `half on the grid', in
the sense that there was no colouring of a $2 \times 2$ square
containing the intersection of a horizontal and a vertical continuous
line.

This situation becomes much more clear by visualisation.
Let us represent the blue coloured $2 \times 2$ squares by their
upper-left corner cells only.
We collect the colourings in the order given by the Maude output of
this experiment. 
A darker blue represents a greater accumulation of colourings on the
respective position.
As we advance through the 831 possibilities, in the order listed by
the output of this experiment, some cells eventually gets darker, but
at an intensity proportional to the accumulation on the respective
positions. 
The first blue colouring is on the left-upper corner of the board.
Since this is the only time we represent a first blue colouring, we
already make it dark.  
Now, let us ignore the first colourings, which for this initial
segment of the experiment happen only at the left-upper corner of the
board, and consider the first 88 results of the experiment for the
second colouring by $B$.  
In the figure below we can see some snapshots of that, representing the
accumulation of the blue colouring at the 23rd, 45th, 67th, and the
89th result, respectively.
At the latter one, the grid is already very apparent.
If we continued this visualisation till the last (i.e. 831st) result,
the grid would pop-up with very sharp clarity.
We can notice some very low accumulation on the positions that do not
correspond to the $3 \times 3$ grid mentioned above, including zero
accumulation on the positions that are not even `half on the grid'.
In this way, the strategy for $B$ emerges clearly as a visual image. 
\begin{figure}[H]
    \centering
    \subfigure{\includegraphics[width=0.16\textwidth]{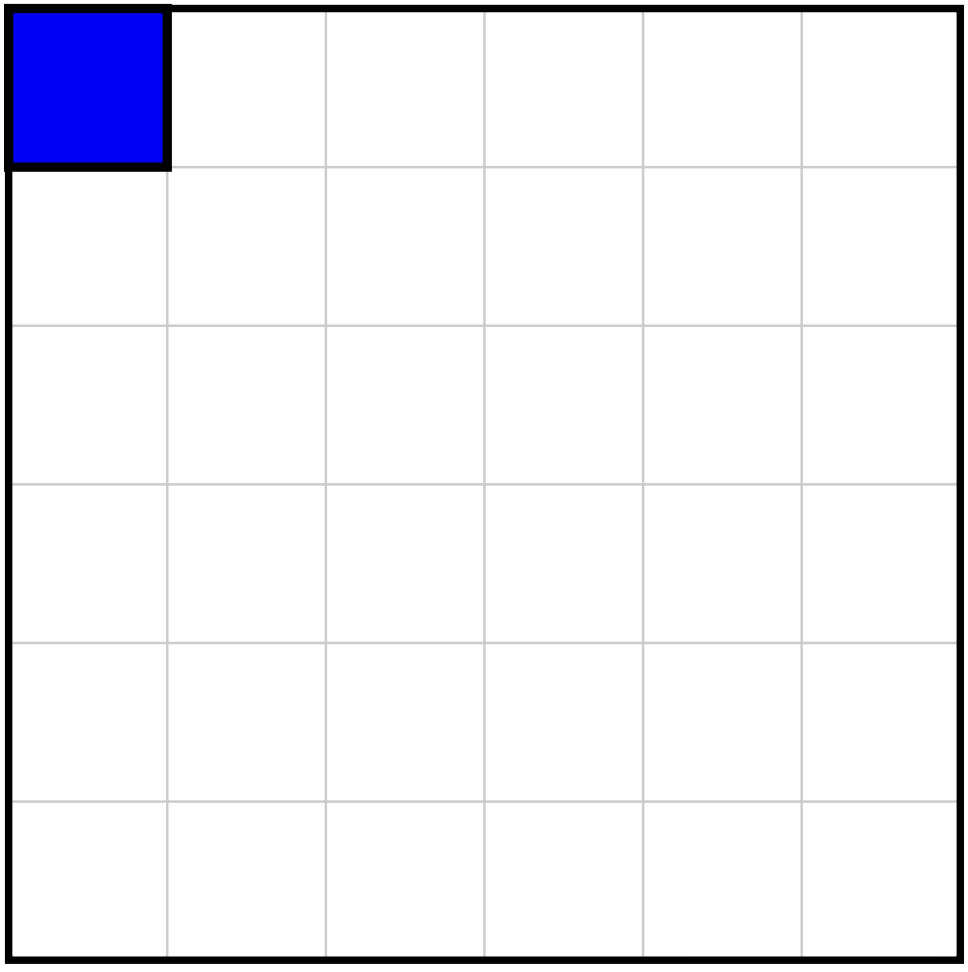}}
    \hspace{1em}
    \subfigure{\includegraphics[width=0.16\textwidth]{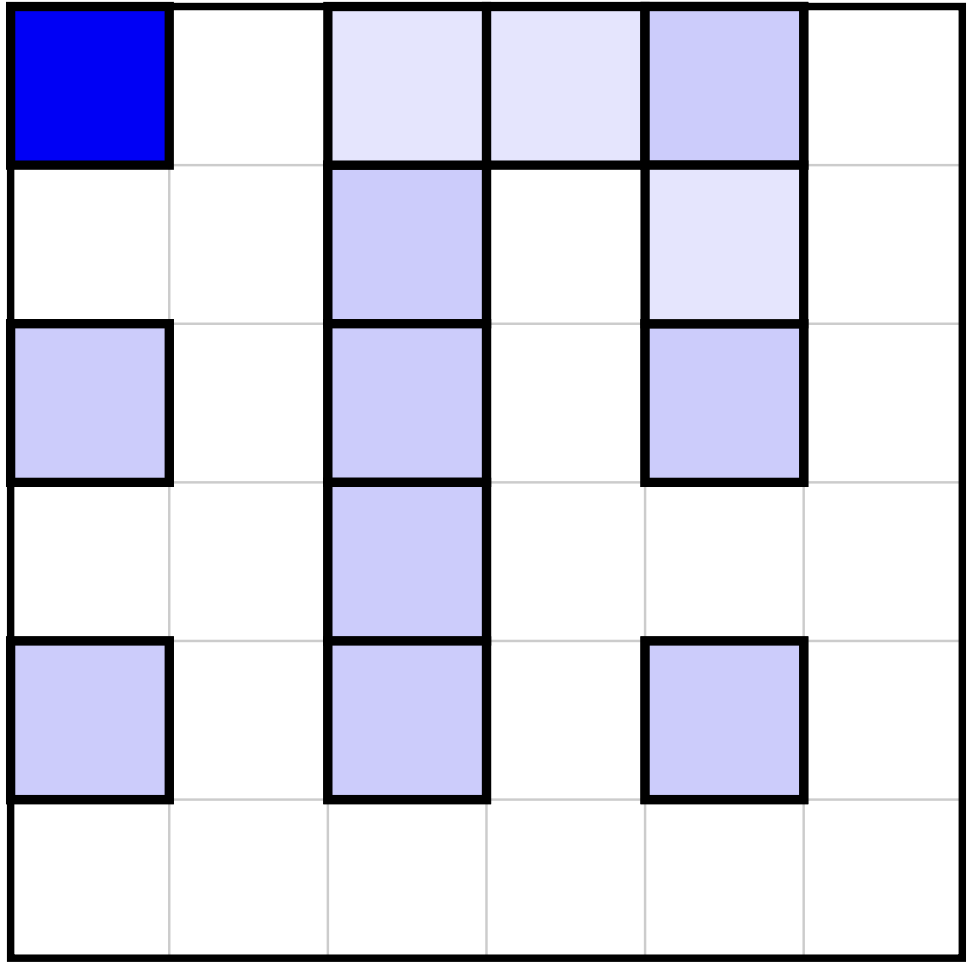}}
    \hspace{1em}
    \subfigure{\includegraphics[width=0.16\textwidth]{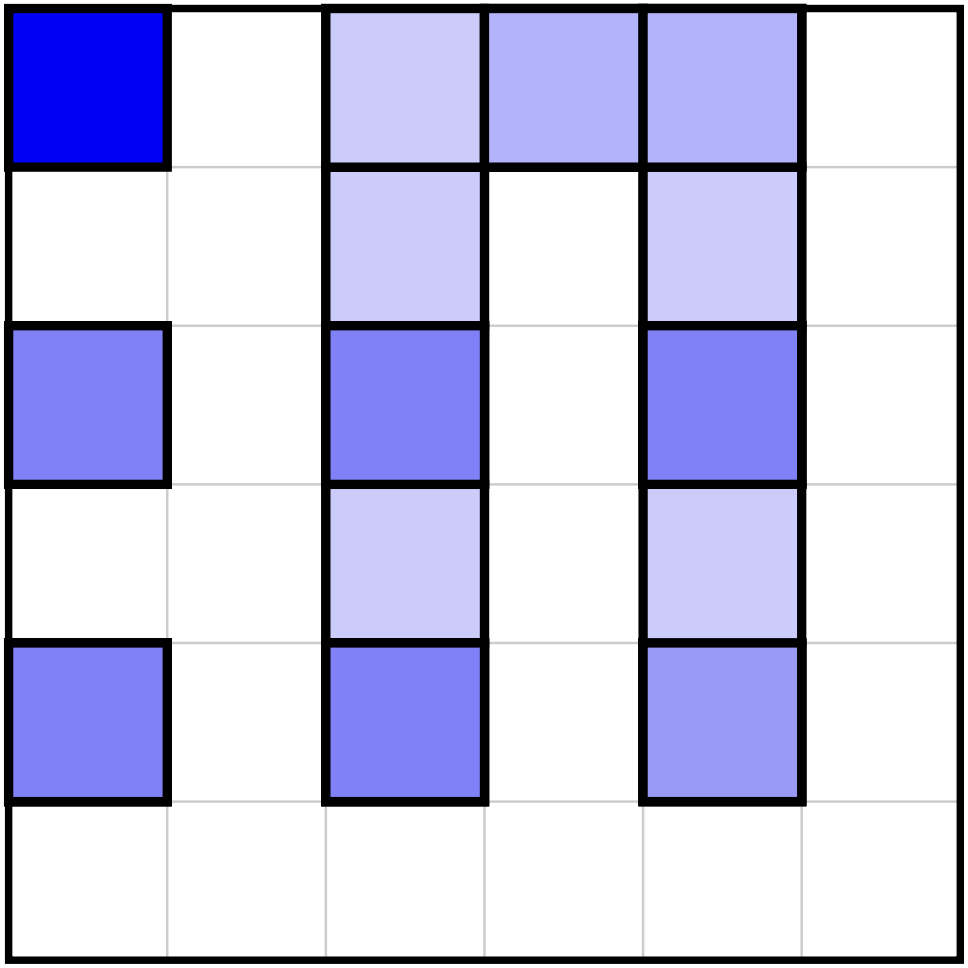}}
    \hspace{1em}
    \subfigure{\includegraphics[width=0.16\textwidth]{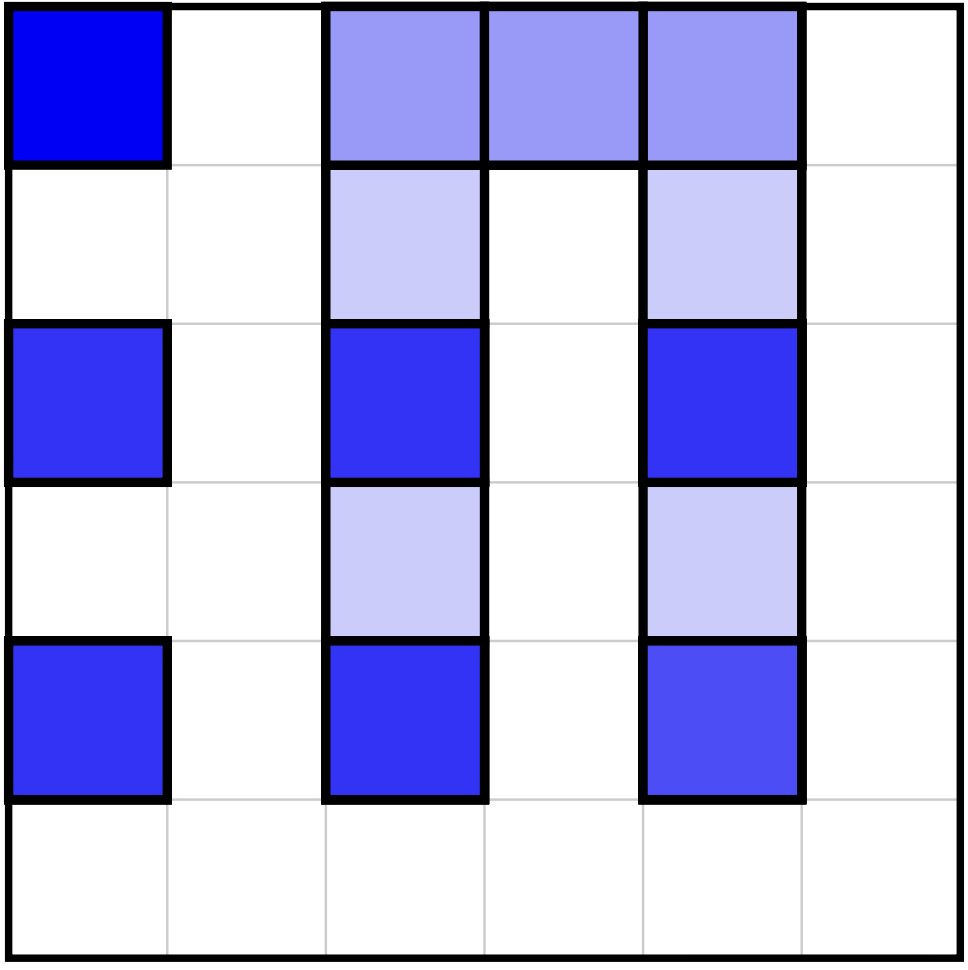}}
    \hspace{1em}
    \subfigure{\includegraphics[width=0.16\textwidth]{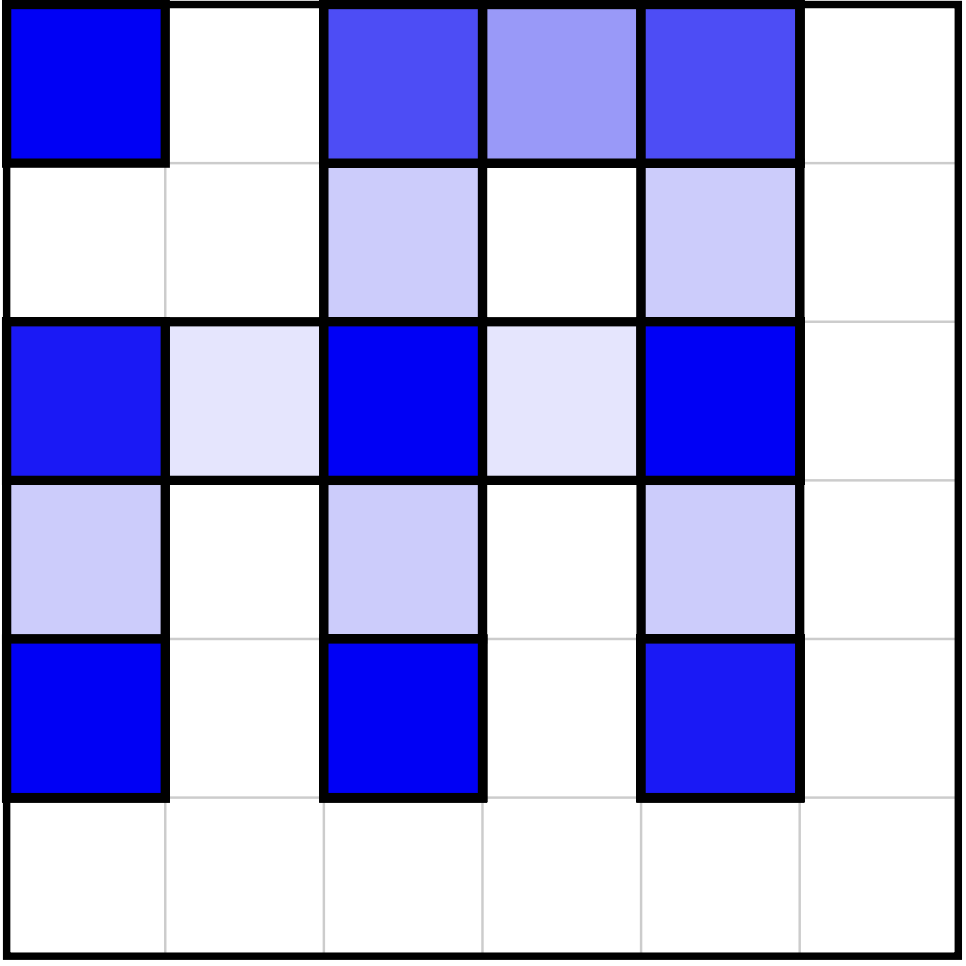}} 
\end{figure}

\paragraph{Understanding the winning strategies.}
From this we can conjecture naturally that $B$'s winning strategy is
to always colour the $2 \times 2$ squares of that grid.
We also consider the fact that, most probably, the $17\%$ of the
cases when the second move was only `half on the grid' is due to a
very poor move by $R$ leading to an (undeserved) wealth of further
winning configurations for $B$. 
This is supported by an analysis of those $17\%$ cases, in \emph{all}
of them $R$'s move was on the border of the first move by
$B$, which does not meet $R$'s own interest to narrow
the set of the positions available for $B$. 
We can have a straightforward argument why that strategy works for
$B$ as follows: 

\begin{prop}
On any $n \times n$ board such that $n$ is even, if $B$ always colours
an available $2 \times 2$ square on the grid like in Figure
\ref{grid2x2}, then he gets at least a draw.
More precisely, if $n = 4k$ he gets a draw and if $n=4k+2$ he can win.

\end{prop}

\begin{proof}
If $B$ manages to colour at least half of the $2 \times 2$
squares on the grid then the blue cells will be at least half of the
$n^2$ cells on the board.
Since $B$ starts to colour and, because at each move $R$
can block at most one new $2 \times 2$ square on the grid, the
conclusion is obvious.
When $n=4k+2$, $B$ can perform $2k^2 + 2k + 1$ moves, which means that
he wins because there will be $4(2k^2 + 2k + 1)$ blue cells.
\end{proof}

\begin{figure}
\begin{center}
\scalebox{1.3}{
\includegraphics{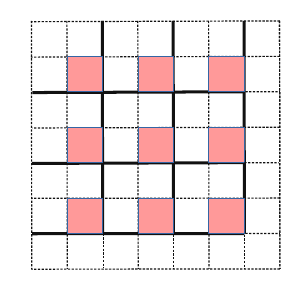}
}
\caption{Candidate positions for red colourings in a winning
  strategy for $R$} 
\label{r-win}
\end{center}
\end{figure}

After learning what is the most efficient strategy for $B$ to
maximise the number of blue cells on the board, we turn to the issue
of the winning strategies for $R$.
We assume a $n \times n$ board such that $n$ is odd.
A maximal grid of $2 \times 2$ squares covers exactly one corner of
the board, so there are four such grids; one of them being shown in
Figure \ref{r-win}. 

Then $R$ should block as many $2 \times 2$ squares on the grid
by colouring into them.
Since this should catch all four grids, the nine red positions in
Figure \ref{r-win} represent the only possibility to achieve that.
This is an informal argument, fully based on blocking the strategy of
$B$ (to maximize the number of the blue cells).
So, it remains to prove that it works.

\begin{prop}
  Let $n = 2k +1$ be an odd positive integer.
  Then a winning strategy for $R$ consists of always colouring
  any available cell with both coordinates even. 
\end{prop}

\begin{proof}
  There are $k^2$ cells with both coordinates even.
  Since with one move, $B$ covers exactly one of these cells,
  at the end $B$ colours $\ceiling{k^2 / 2}$ cells, hence there
  will be $4 \ceiling{k^2 / 2}$ blue cells.
  Because
  \[
  8  \ceiling{k^2 / 2} \leq 8 (k^2 / 2 + 1/2) = 4k^2 + 4 <
  4k^2 + 4k +1 = (2k+1)^2 = n^2
  \]
  we end up with at least as many red as blue cells.
  Since $n$ is odd, this obviously means more red than blue cells.
\end{proof}

The conclusion of this case study is that by experimenting with
running through the winning strategies of the opponent of $R$,
we understood the brute winning strategies for $R$ herself,
which led to the formulation of an adequate conjecture and further on
to its proof. 

\subsection{Computing traces of winning strategies by equational
  programming}

We can obtain somehow
data similar to that obtained through the method
introduced in Section \ref{poa-sec}, but this time by equational
programming. 
This method relies on upgrading the module \texttt{WINS} by replacing
the data type of the Booleans with a theory of `traces'.
This can be instantiating to whatever concrete data type of traces we
would need, such as sets of moves, or lists, or even trees, etc.
The main points of this upgrade are as follows:
\begin{itemize}

\item The Boolean constant \texttt{true} unfolds to the set of
  possible traces.

\item The Boolean constant \texttt{false} is now represented by an
  element $0$ that is added to the set of the traces, itself not being
  considered as a trace.

\item The value of $\texttt{w}(p, c)$ is either $0$ (there is no winning
  strategy for $p$ at configuration $c$ or a trace
  [of a winning strategy or of a set of winning strategies]). 

\item There is an aggregation operation on traces, denoted
  \texttt{\_\_}.
  For instance, if traces are lists (of moves) then this would be
  interpreted as list concatenation, if they are sets then the
  aggregation may be union.
  The aggregation with $0$ `kills' any trace.

\item The configurations should carry some information about the moves
  that lead to that particular configuration. 
  At the abstract level of \texttt{CONFIG} this would come as a new
  operation:
  \begin{maude}
  op move : Config -> Trace .  
  \end{maude}
  
\end{itemize}
Below is the specification of a generic module for traces:
\begin{maude}
fth TRACES is
  sorts Trace Trace+ .
  subsort Trace < Trace+ . 
  var t : Trace+ .
  vars t1 t2 : Trace .
  op 0 : -> Trace+ .  
  op e : -> Trace .   --- empty trace
  op __ : Trace Trace -> Trace [assoc id: e] .
  op __ : Trace+ Trace+ -> Trace+ [assoc id: e] . 
  eq 0 t = 0 .
  eq t 0 = 0 .
endfth
\end{maude}
Then we can have the following upgrade for \texttt{WINS}, for tracing
the winning strategies of $R$:
\begin{maude}
fmod WINS-T{X :: GAME-TREE-T} is
  op w : X$Player X$Config -> X$Trace+ .
  op w-aux : X$Player X$SetConfig -> X$Trace+ .

  var p : X$Player . 
  var c : X$Config .
  var S : X$SetConfig .

  eq w(p, c) = if halted(p, c)
                  then if wonR(p, c) then e else 0 fi
                  else w-aux(opponent(p), next-config(p, c))
	       fi .
  eq w-aux(R, empty) = e .
  eq w-aux(R, (c, S)) = w(R, c) w-aux(R, S) .
  eq w-aux(B, empty) = 0 .
  eq w-aux(B, (c, S)) = if w(B, c) == 0 then w-aux(B, S)
                                        else (w(B, c) move(c))
		        fi .  
endfm
\end{maude}
In general, this tracing method can be used instead of the
straightforward use of \texttt{WINS}, whenever we need traces.

\paragraph{How it works for the board game.}
In this case we have to do the following upgrades to the instantiation
that we discussed in the Section \ref{concrete-sec}:
\begin{itemize}

\item We consider a trace to be the set of the positions that
  represent the totality of the moves performed by $R$ in a
  winning strategy, relative to a given configuration.
  
\item We add a fourth field to the configurations which are now
  represented as a 4-tuple $(P, P_2, k, M)$ where $M$ stands for the
  last move leading to $P$; $M$ is a position on the board.

\item $\texttt{move}(P, P_2, k, M) = M$.

\end{itemize}
The full code for this experimentation can be found in repository
\cite{board-game}.

Like with the previous method, by computing traces for $R$ we
do not get data that is helpful for understanding a smart winning
strategy, so we may turn to computing `winning' strategies for
$B$ instead.
By doing so, we were able to note that $B$ always colour at least
`half on the grid', and, moreover, at least half of the trace was on
the grid. 
This is a hint to the smart winning strategy for $B$, very similar but
weaker than what we got by the non-deterministic rewriting method.
We believe that in other situations this kind of tracing may prove
stronger, and provide even more substantial information that can be
used to conjecture smart strategies.  

\paragraph{Further remarks on using \texttt{TRACES}.}
In order to simplify our presentation we did here a straightforward
version of \texttt{WINS-T}.
Its instance for the board game runs at much lower speed than the
corresponding instance of \texttt{WINS}.
However, in repository \cite{more-generic-wins} there is a  more
sophisticated version of \texttt{WINS-T} (called \texttt{TRACE-WINS})
that runs at approximately the same speed with \texttt{WINS}.
This owes to a generalisation of the operations \texttt{and-then} and
\texttt{or-else} from Booleans to traces.
Moreover, this upgraded version of \texttt{WINS-T} comes as an
instance of a more abstract specification that generalises both
\texttt{WINS} and \texttt{WINS-T}.

\paragraph{Computing traces versus computing with transitions.}
There are significant differences between the experimental methods of
this section and of Section \ref{poa-sec}.
In the latter case, the result albeit partial in depth is
comprehensive in breadth (all relevant sequences of moves up to a
certain bound are explored).
The sequences of moves can be seen clearly.
The partiality in depth and the slow speed owe to the size of the
\POA-model and to the fact that each transition/rule application
involves an evaluation of the function \texttt{w}.
Since these evaluations happen for boards with only a few missing
cells, they are costly.
Moreover, there is an aspect of redundancy, the same sequence of
$B$-moves appears multiple times because there are many possible
$R$-moves between them.
On the other hand, the traces method is faster (especially in the
`smarter' version mentioned above) because it computes much less of
\texttt{w}, but the output is less comprehensive.
In the traces-as-sets version the output is also more difficult to
interpret because the sequentiality is missing.
These considerations refer to the particular case of the bord game,
but they also carry an element of generality. 

\section{Conclusions and Future Research}

Based on a concept of strategy as subtree (of the game tree), in this
paper we have developed our own version of Zermelo's theorem about
strategies in combinatorial games.
This allows for a smooth equational logic-based generic computational
modelling in which the game trees of concrete games appear as its
algebras.
These game trees can be specified equationally too, thus leading to
executable concrete instances of the generic specification. 
By running these programs we can establish the existence of brute
strategies of particular concrete games.
This led us to a form of experimental mathematics, further supported
by some additional methods, that in some cases may provide crucial
insight supporting the smart formulation of strategies for general
concrete games, and finally to mathematical proofs validating the
strategies. 
In addition to that we have also discussed an example from computing
science were just establishing the existence of winning strategies
serves the purpose.
Moreover, that example involved a double-stage parameterisation. 

This work opens up at least two important avenues for further research
and development.
\begin{enumerate}[leftmargin=1.5em]

\item We believe that other types of games, such as other
  extensive-form games, might be suitable candidates for the type of 
  logic-based computational modelling that we introduced in this
  paper. 
  We would like to explore such possibilities.
  
\item Overall, for our purpose, Maude has been very helpful as it
  inherits the powerful OBJ3 equational logic-based specification and
  programming paradigm quite faithfully, and, on top of this, its
  rewrite engine is unequalled in terms of power and sophistication.
  Moreover, Maude extends the OBJ paradigm with non-deterministic
  rewriting, which is very useful for symbolic experimental
  mathematics in combinatorics, but not only.
  However, there are some aspects of Maude that need to be reformed to
  serve our experimental mathematics purpose better.
  \begin{itemize}

  \item The module system can be made to be more permissive, in the
    direction of the module systems of OBJ3 and CafeOBJ.
    This is facilitated by the $\POA$ semantics, which allows for the
    general modularisation technology provided by the
    institution-theoretic approach. 

  \item $\POA$ semantics also allows for an equal (like in CafeOBJ)
    rather than hierarchical (like in Maude) treatment of equations
    and transitions.
    This would increase the specification power by allowing equations
    conditioned by transitions, via CafeOBJ-styled built-in predicates
    \texttt{==>}. 

  \item The style of using transitions that we employed in Section
    \ref{poa-sec} has an imperative object-oriented flavour but
    involves some tedious duplications.
    This can be solved by introducing a new smarter dedicated syntax. 

  \end{itemize}
  The former two proposed upgrades clash with some feature of Maude as
  it is now (we do not discuss details here). 
  Therefore, the best way is to develop a new dedicated language
  around a simple fragment of Maude, also employing the Maude
  rewriting engine.
  We believe that this can be achieved smoothly within the SpeX
  environment \cite{spex}.

  \item Investigate possibilities to have a proper generic
    specification of the two transitions of Section \ref{poa-sec}. 

\end{enumerate}

\paragraph{Acknowledgement.}
I am grateful to Tashi Diaconescu for encouragement and appreciation,
and for helping with various issues related to this work.
First, he provided the heaps and the board games and the respective
problems. 
This raised my interest in computational experiments with these
problems, which led to the computational modelling presented here.
We also discussed a lot of technical and presentation details of this
work. 
As I developed Theorem \ref{strategy-thm1} on my own without any
awareness of Zermelo's work, Tashi showed me Zermelo's result.
He also greatly helped with \LaTeX\ issues when writing down this
paper.
I am also grateful to my colleague Ionu\c{t} \c{T}u\c{t}u for
developing the code for the instantiation to the bisimilarity game,
for writing the code unifying in abstraction \texttt{WINS} and
\texttt{WINS-T}, and also for helping with issues that arose while
working with Maude.  
Both reviewers have been great contributors to this paper through
their constructive suggestions about several details that were not
right in the original draft, and about extending the work with more
applications and more explanations.

\bibliographystyle{abbrv}   
\bibliography{/Users/diacon/TEX/tex}

\appendix
\section{Generating the game trees for the heaps game}
\label{heaps-appendix}

First, the players.
This essentially the same specification like in the parameter, but
this time with initial semantics, which means that now the players are
only $B$ and $R$.
\begin{maude}
fmod PLAYERS-HP is
  sort PlayerHP .
  ops B R : -> Player .
  op opponent : Player -> Player .
  eq opponent(B) = R .
  eq opponent(R) = B .
endfm
\end{maude}

The configurations are two heaps of stones, each heap is
represented by a non-negative integer number that gives the number of
the stones in the respective heap.
This is the only information that is relevant for playing the game.
Below we specify this as a data type.
\begin{maude}
fmod TWO-HEAPS is
  protecting INT .
  sort TwoHeaps .
  op (__) : Int Int -> TwoHeaps .
\end{maude}
The terminal configurations are the pairs of empty heaps; this
situation is specified by the definition of \texttt{halted} (the
corresponding equation).
\begin{maude}
fmod HALTED-HEAPS is
  protecting BOOL .
  protecting PLAYERS-HP .
  protecting TWO-HEAPS . 
  op halted : Player TwoHeaps -> Bool .
  vars k n : Int .
  var p : Player .
  eq halted(p, (k n)) = ( (k n) == (0 0) ) .
endfm
\end{maude}
For this particular game, we specify the concrete criteria when $R$ is
a winner when the game halted.
According to the specification of the game, this happens when Benny
has to move but he cannot. 
\begin{maude}
fmod WON-HEAPS is
  protecting BOOL .
  protecting HALTED-HEAPS .
  op won : Player TwoHeaps -> Bool .
  var p : Player .
  var c : TwoHeaps . 
  eq won(p, c) = (B == p) and halted(p, c) . 
endfm
\end{maude}

Now we move towards a concrete specification for \texttt{next-config},
the core of the instantiation process. 
As configurations are `pairs of heaps', sets of configurations are
sets of \texttt{TwoHeaps}.
For this we use the generic predefined module doing sets,
i.e. \texttt{SET\{X :: TRIV\}} and instantiate its parameter and obtain
\texttt{SET\{TwoHeaps\}}. 
\begin{maude}
view TwoHeaps from TRIV to TWO-HEAPS is
  sort Elt to TwoHeaps .
endv
\end{maude}

For specifying the moves in the game between Benny and Rebecca, we
need the auxiliary operation \texttt{sub} that subtracts $m$ from the 
second  heap as many times as possible and it returns all possible
results as a set of \texttt{TwoHeaps}.
That we subtract only form the second heap is based on the following
arrangement that will simplify and speed up the execution of the
program. 
Because in the problem the order of the heaps is immaterial, a pair of
heaps, which is a configuration in the game, can be considered
ordered, i.e. we arrange the pairs of heaps $(a \ b)$ such that
$a \leq b$. 
\begin{maude}
fmod ORDER-SET-OF-TWO-HEAPS is
  protecting SET{TwoHeaps} .
  op order : Set{TwoHeaps} -> Set{TwoHeaps} . 
  vars a b : Int .
  var S : Set{TwoHeaps} .
  ceq order(((a b), S)) = ((a b), order(S)) if a <= b .
  ceq order(((a b), S)) = ((b a), order(S)) if b < a .
  eq order(empty) = empty .
endfm
\end{maude}

\begin{maude}
fmod SUBTRACT-FROM-TWO-HEAPS is
  protecting SET{TwoHeaps} .
  op sub : TwoHeaps Int -> Set{TwoHeaps} .
  vars a b m : Int .
  ceq sub((a b), m) = (a (b + (- m))), sub((a (b + (- m))), m)
      if m <= b .
  ceq sub((a b), m) = empty if b < m .
  endfm
\end{maude}
Now we can compute the next pairs of heaps (\texttt{next-heaps} below)
obtained after a move in the game.
\begin{maude}
fmod NEXT-HEAPS is
  protecting ORDER-SET-OF-TWO-HEAPS .
  protecting SUBTRACT-FROM-TWO-HEAPS .
  op next-heaps : TwoHeaps -> Set{TwoHeaps} .
  vars a b m : Int .
  var S : Set{TwoHeaps} .
  eq next-heaps (0 0) = empty .
  eq next-heaps (0 a) = (0 0) .
  ceq next-heaps (a a) = (0 a) if 0 =/= a .
  ceq next-heaps (a b) = (0 a), (0 b), order(sub((a b), a))
      if (0 =/= a) and (a < b) .
endfm
\end{maude}

Finally, for this particular game, we build the instance of the
abstract parameter \texttt{GAME-TREE}: 
\begin{maude}
fmod HEAPS-GAME is
  protecting HALTED-HEAPS .
  protecting WON-HEAPS .
  protecting NEXT-HEAPS . 
endfm 
\end{maude}

\end{document}